\documentclass[lettersize,journal]{IEEEtran}
\usepackage{amsmath,amsfonts}
\usepackage{algorithmic}
\usepackage{algorithm}
\usepackage{array}
\usepackage[caption=false,font=normalsize,labelfont=sf,textfont=sf]{subfig}
\usepackage{textcomp}
\usepackage{stfloats}
\usepackage{url}
\usepackage{verbatim}
\usepackage{graphicx}
\usepackage{cite}
\usepackage{multirow}
\usepackage{booktabs}
\usepackage{array}
\usepackage{float}
\usepackage{graphicx} % for \rotatebox
\usepackage{adjustbox}
\usepackage{multirow}
\usepackage{array} % 提供 m{..} 与 \arraybackslash
\usepackage{color}
\newcolumntype{M}[1]{>{\centering\arraybackslash}m{#1}}
\renewcommand{\textcolor}[2]{#2}

\begin{document}

\title{Hierarchical Semantic Learning for Multimodal Sentiment Analysis with Missing Modalities}

\author{Chunlei Meng,~\IEEEmembership{Student Member,~IEEE}, clmeng23@m.fudan.edu.cn, \\Rong Fu, Zhan Chen, Yi Liu,~\IEEEmembership{Graduate Student Member,~IEEE}, \\ Zhuo Zou,~\IEEEmembership{Senior Member,~IEEE}, LiDa Xu,~\IEEEmembership{Fellow,~IEEE}, Chun Ouyang*,~\IEEEmembership{Member,~IEEE}, and Zhongxue Gan\\ \textbf{Accepted for Publication in IEEE Transactions on Multimedia (TMM), 2026.}

\thanks{C. Meng, Y. Liu, Z. Gan, and C. Ouyang are with the College of Intelligent Robotics and Advanced Manufacturing, Fudan University, Shanghai, China. R. Fu is with the University of Macau. Z. Chen is with the Aerospace Information Research Institute, Chinese Academy of Sciences. Z. Zou is with the School of Information Science and Technology, Fudan University. L. Xu is with the Department of Information Technology, Old Dominion University. \\ *Corresponding Author: Chun Ouyang.}}

% <-this % stops a space


        % <-this % stops a space



\maketitle

\begin{abstract}
Multimodal Sentiment Analysis (MSA) is increasingly deployed in real-world multimedia systems, yet practical inputs often exhibit unpredictable modality missingness due to sensor failures, privacy masking, or unstable transmission. \textcolor{red}{Existing approaches for MSA under missing modalities either learn reliability-aware representations or reconstruct missing inputs, and most of them operate on a single flattened encoder output.} This flattened modeling ignores the inherent semantic hierarchy of deep encoders, as well as the semantic interactions within and between modalities, leading to semantic averaging and ill-posed mapping, which in turn limit performance. To address this limitation, this study proposes a novel \textbf{Hierarchical Semantic Learning} framework (HSL) that explicitly exploits semantic hierarchy for robust inference with missing modalities. HSL introduces \textbf{Hierarchical Semantic Decoupling} to expose low-level, mid-level, and high-level feature streams for each modality encoder. It further presents \textbf{Cross-Modal Cross-Level Semantic Reconstruction} to perform ladder-style reconstruction, where high-level semantics from observed modalities act as global anchors and progressively guide the recovery of mid-level structure and low-level details for missing modalities. To stabilize reconstruction across levels, \textbf{Hierarchical Cascade Error Supervisor} regularizes both observed and reconstructed features via prototypical contrastive learning to enforce level-specific semantic consistency and to reduce error propagation. Finally, \textbf{Hierarchical Interactive Fusion} models dense interactions across modalities and semantic levels to produce stable predictions under diverse missingness patterns. Extensive experiments on CMU-MOSI, CMU-MOSEI, IEMOCAP, and CH-SIMS under various missing-modality settings show that HSL \textcolor{red}{outperforms the baselines in most settings}, validating the effectiveness of hierarchical semantic learning for MSA with missing modalities.
\end{abstract}

% Extensive experimental results show that HSL outperforms existing models in various missing modalities and achieves state-of-the-art performance on multiple benchmarks (MOSI:0.965, MOSEI:0.635, SIMS:0.472, IEMOCAP-FOUR:74.36, IEMOCAP-SIX:56.05), validating the effectiveness of hierarchical semantic learning for MSA with Missing Modalities.


\begin{figure}[htbp]
    \centering
    \includegraphics[width=\linewidth]{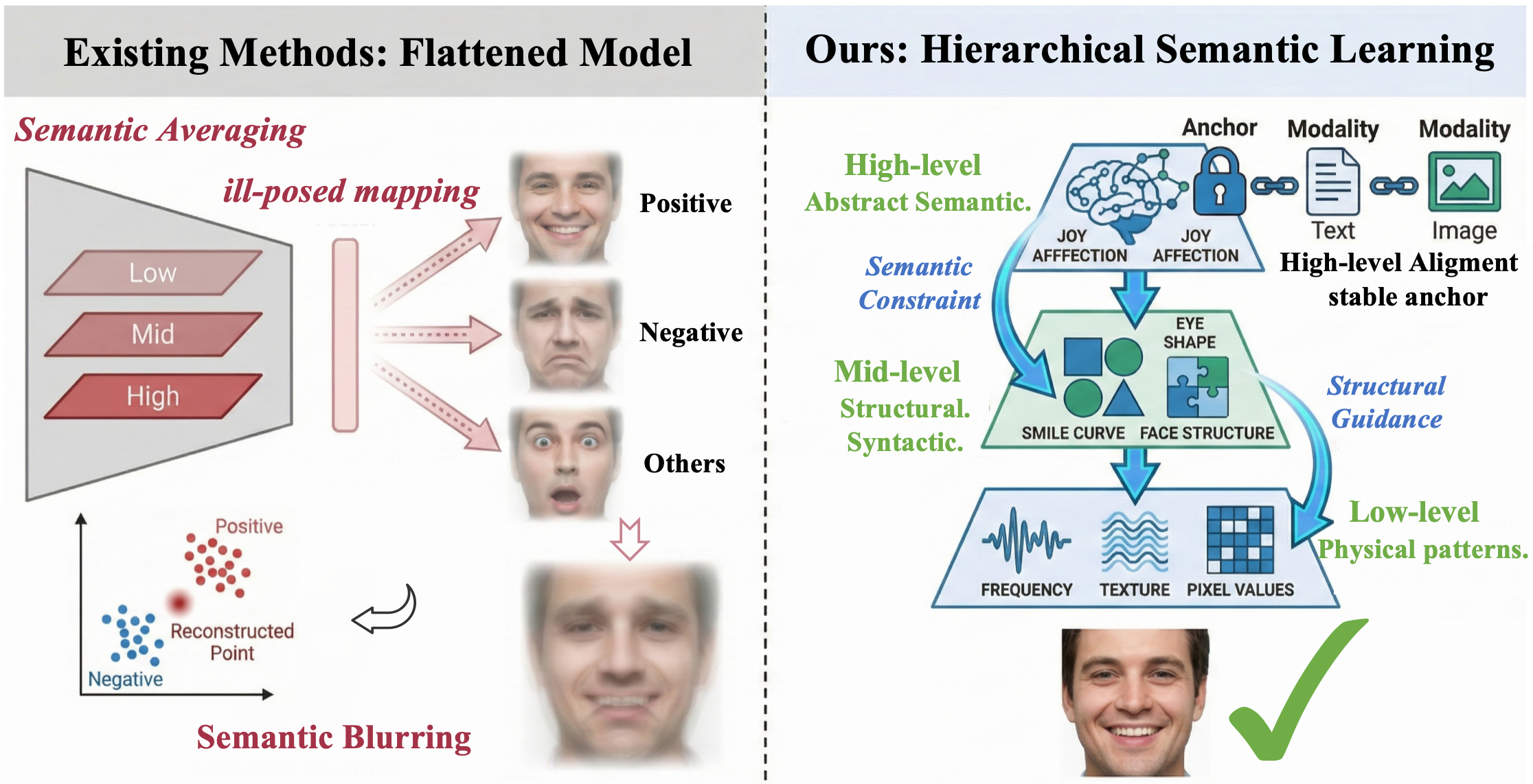}
    \caption{Motivation for HSL. (left) Baseline: Flattened outputs disrupt semantic hierarchy, leading to ill-posed mapping and representation blurring in missing cases. (right) HSL: Leverages hierarchical complementarity via anchor-guided top-down reconstruction, ensuring robust and stable representations.}
    \label{fig:motivation}
\end{figure}

\section{Introduction}
\IEEEPARstart{M}{ultimodal} Sentiment Analysis (MSA) aims to infer human affect by jointly modeling heterogeneous cues such as language, acoustic signals, and visual expressions~\cite{TSD,GCL,PSM-WSM,MRCF}. Although multimodal cues usually improve performance when inputs are complete, real-world deployments often face modality-level and intra-modal missingness due to sensor failures, privacy masking, or unstable transmission~\cite{ROSA,MLCL,DAR,LNLN,P-RMF}. Maintaining reliable performance across diverse missingness patterns is therefore a key requirement for practical MSA systems~\cite{NIAT,HKD-MER,IMDer,GCNet}.

Existing robust MSA methods are commonly grouped into two paradigms. The first paradigm improves robustness via reliability-aware representation learning and fusion. For example, ROSA~\cite{ROSA} estimates modality-wise reliability under uncertain missing patterns and adaptively reweights fusion. MLCL~\cite{MLCL} introduces multi-level contrastive objectives to encourage consistent unimodal and multimodal representations under partial corruption. HKD-MER~\cite{HKD-MER} transfers complementary affective cues via hierarchical knowledge distillation from a full-modality teacher to partially observed students. NIAT~\cite{NIAT} strengthens robustness through noise imitation and adversarial training. Disentanglement-based methods such as MISA~\cite{misa} and FDMER~\cite{FDMER} also learn structured representations by separating modality-invariant and modality-specific factors. The second paradigm explicitly reconstructs missing information. DAR~\cite{DAR} reconstructs missing features in a factorized space after decoupling sentiment-relevant components, while P-RMF~\cite{P-RMF} performs proxy-driven generation in a proxy space designed for stable inference under incompleteness.

% LNLN~\cite{LNLN} further optimizes over multiple missingness patterns to learn representations robust to partial observations.

% , and EAU~\cite{EAU} further modulates fusion by unimodal aleatoric uncertainty

Despite this progress, most existing methods enforce robustness on a single flattened representation at the encoder output. They either regularize the final embedding to be noise tolerant~\cite{ROSA,MLCL,EAU}, or impute the missing-modality embedding in a single-stage manner~\cite{DAR,P-RMF,LNLN}. \textcolor{red}{Only a few methods, such as EMT~\cite{EMT} and HRLF~\cite{HRLF}, exploit more than one representation level. However, incorporating multiple representation levels does not necessarily establish explicit dependencies across encoder depths during missing-modality reconstruction. In particular, how inferred high-level semantics from observed modalities can progressively guide the reconstruction of lower-level features remains underexplored. This distinction matters because deep encoders are inherently hierarchical. Low-level layers capture fine-grained physical patterns, mid-level layers encode intermediate structure and composition, and high-level layers summarize more abstract semantics~\cite{2019BERT}.} Flattening this hierarchy makes missing-modality imputation more underdetermined, because similar high-level semantics can correspond to multiple plausible mid-level structures and low-level realizations~\cite{SRGAN}. Under Euclidean reconstruction losses (e.g., $\ell_2$/MSE), minimizing the expected error encourages predictions to approach the conditional mean, a behavior that often \textcolor{red}{produces} overly smoothed outputs in conditional generation~\cite{SRGAN,pix2pix}. In MSA, this averaging effect pushes reconstructed features toward regions that are common across sentiment classes, causing class-specific cues to fade and class boundaries to overlap. Consequently, the separation margin between sentiment categories becomes smaller, leading to weaker fine-grained sentiment discrimination~\cite{UniMF,TagEnc}.

Moreover, cross-modal dependencies vary across semantic levels. High-level semantics are often more alignable across modalities, mid-level structures share only partial correspondence, and low-level physical cues exhibit weaker and noisier correlations. Forcing a single embedding to absorb these heterogeneous correlations can lead to ill-posed mapping, which amplifies the distribution mismatch between reconstructed features and true features, especially for low- and mid-level cues~\cite{P-RMF,SRGAN}. Fig.~\ref{fig:motivation} illustrates these limitations, flat single-stage modeling is insufficient under severe and uncertain missingness. Robustness should instead be established through a level-specific paradigm that respects semantic hierarchy, where high-level semantics provide stable constraints and mid-level and low-level cues are progressively refined~\cite{MissModal,TagEnc}.

\begin{figure*}[htbp]
\centering
\includegraphics[width=1\linewidth]{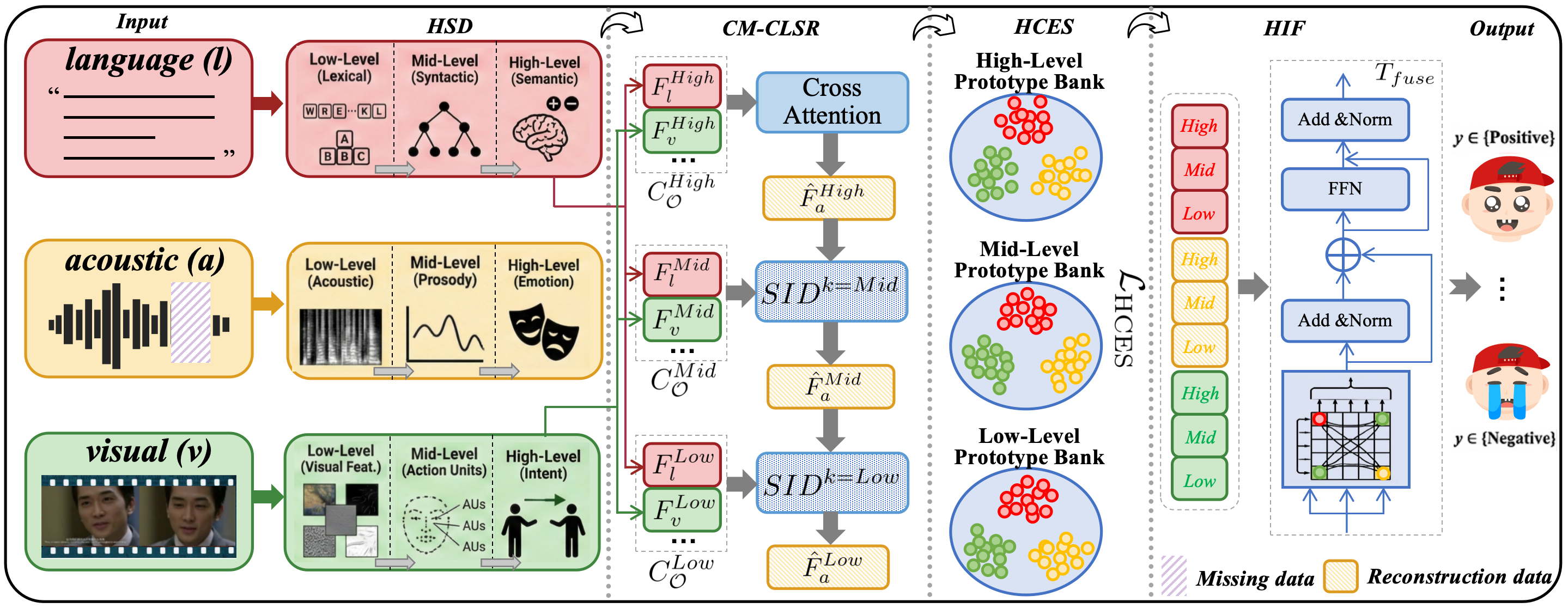}
\caption{Overview of the Hierarchical Semantic Learning (HSL) framework, \textcolor{red}{illustrated with a missing acoustic modality.} HSL consists of four components: \textbf{HSD} decomposes each observed modality into low-, mid-, and high-level semantics; \textbf{CM-CLSR} reconstructs missing mid-/low-level features from high-level cues via the Semantic Injection Decoder (SID); \textbf{HCES} regularizes level-wise reconstructions using a contrastive loss; and \textbf{HIF} fuses all hierarchical features with a Transformer to produce the final prediction. Solid arrows indicate data flow, and dashed arrows indicate supervision.}
\label{fig:framework}
\end{figure*}

Motivated by this hierarchy-induced underdetermination, this study proposes a \textbf{Hierarchical Semantic Learning} framework (HSL) for robust MSA with missing modalities (Fig.~\ref{fig:framework}). HSL restructures each modality encoder into an explicit three-level hierarchy and couples it with anchor-guided top-down reconstruction and hierarchical fusion. Specifically, \textbf{Hierarchical Semantic Decoupling (HSD)} exposes low-level, mid-level, and high-level feature streams to preserve semantic progression. \textbf{Cross-Modal Cross-Level Semantic Reconstruction (CM-CLSR)} performs ladder-style reconstruction, using high-level semantics from observed modalities as anchors and propagating information downward to refine mid-level structures and low-level details. To mitigate error accumulation and overly smoothed reconstructions, \textbf{Hierarchical Cascade Error Supervisor (HCES)} introduces level-specific prototypical contrastive learning to enforce semantic consistency for both observed and reconstructed features. Finally, \textbf{Hierarchical Interactive Fusion (HIF)} models dense interactions across semantic levels and modalities, enabling high-level semantics to route and refine lower-level cues for stable prediction under diverse missingness patterns. Our contributions are summarized as follows:
\begin{itemize}
    \item \textcolor{red}{A level-specific framework, HSL, is proposed that focuses on cross-modal, top-down reconstruction across encoder depths: the inferred high-level representation conditions the reconstruction of mid- and low-level features, with observed modalities providing evidence at each corresponding level.}
    \item HSD and CM-CLSR are designed to preserve level-wise feature streams and to perform ladder-style reconstruction across high-, mid-, and low-level semantics with semantic constraints.
    \item HCES and HIF are introduced to enforce level-wise semantic consistency and to model cross-modal and cross-level interactions, yielding stable performance across diverse missingness patterns.
\end{itemize}

\section{Related Works}

\subsection{Multimodal Sentiment Analysis} 
MSA has evolved from basic fusion architectures to sophisticated representation learning~\cite{AHRA,CLCR,DBR}. Early works such as TFN~\cite{TFN} and LMF~\cite{LMF} focused on modeling high-order interactions, while Transformer-based methods like MulT~\cite{MulT} leveraged cross-modal attention for unaligned sequence modeling. Despite their success, these models typically assume full modality availability during inference. To enhance robustness, subsequent research transitioned toward structured representation learning. Methods like MISA~\cite{misa}, Self-MM~\cite{self-mm}, and FDMER~\cite{FDMER} utilize disentanglement, self-supervised objectives, or feature decomposition to decouple modality-invariant and specific factors. However, these representations are primarily derived from single-stage encoder outputs. They lack an explicit internal semantic hierarchy, which limits their effectiveness with severe modality missingness. Although previous work TSDA~\cite{TSDA} and HSL are both developed for MSA, they address different robustness assumptions. TSDA~\cite{TSDA} improves representation learning in complete-modality scenarios via temporal-spatial decoupling, while HSL targets missing-modality inference by restoring low-, mid-, and high-level semantic evidence from observed modalities. Thus, HSL contributes a hierarchical semantic restoration framework for incomplete inputs, which is conceptually distinct from temporal-spatial factor decoupling.

\subsection{Multimodal Learning with Missing Modalities}

Multimodal learning aims to mitigate performance degradation under missing or degraded modalities. Current research primarily follows three trajectories. First, reliability-driven fusion suppresses unreliable or ambiguous inputs by modeling modality-wise importance or aleatoric uncertainty (e.g., ROSA~\cite{ROSA}, EAU~\cite{EAU}). Second, regularization-based methods enhance robustness through auxiliary learning objectives; for instance, MLCL~\cite{MLCL} and MIG-HCL~\cite{MIG-HCL} utilize contrastive learning for representation consistency, while HKD-MER~\cite{HKD-MER} and CorrKD~\cite{CorrKD} employ knowledge distillation to transfer information from full-modality teachers to incomplete students. Additionally, NIAT~\cite{NIAT} leverages adversarial training to counter modality perturbations. Third, generative and unified learning explicitly addresses incompleteness by joint optimization across missing patterns (LNLN)~\cite{LNLN} or feature reconstruction. Specifically, DAR~\cite{DAR} decouples sentiment factors for factorized reconstruction, and P-RMF~\cite{P-RMF} performs proxy-driven generation in a stabilized latent space. 

\textcolor{red}{Several studies have explored related directions for incomplete multimodal learning. EMT~\cite{EMT} combines low-level feature reconstruction with high-level representation attraction between complete and incomplete views. HRLF~\cite{HRLF} learns hierarchical representations through cross-modal translation, multi-scale mutual information maximization, and hierarchical adversarial alignment. DPCN~\cite{DPCN} employs class-specific prototypes of missing modalities and sample-to-prototype affinities to complete missing features in decoupled shared and modality-specific spaces. Beyond multimodal learning, Ladder Networks~\cite{LadderNetworks} perform layer-wise denoising reconstruction through lateral connections between encoder and decoder representations. These studies establish important precedents for hierarchical representation learning, prototype-based completion, and layer-wise reconstruction. HSL differs in its reconstruction dependency: the high-level representation of a missing-modality is first inferred from observed modalities and then serves as the conditioning query for sequential mid- and low-level reconstruction, with same-level observed features providing complementary evidence. Thus, HSL focuses on cross-modal, top-down reconstruction for incomplete MSA rather than introducing these individual components independently.}

\textcolor{red}{Although EMT~\cite{EMT} and HRLF~\cite{HRLF} exploit multiple representation levels, they do not explicitly use the inferred high-level representation of a missing-modality to condition the sequential reconstruction of mid- and low-level features.} Consequently, these approaches struggle to maintain cross-level consistency when error propagation occurs under severe missingness. To address this, we propose HSL, which shifts the focus from flat feature alignment to hierarchical structural restoration. Unlike prior works, HSL leverages high-level semantics as anchors to guide a top-down reconstruction process, ensuring that the entire encoding hierarchy remains stable and complementary even when specific modalities are absent.

\section{Method}

\subsection{Problem Formulation}
\label{sec:III-A}
Let $\mathcal{M}=\{\mathrm{l},\mathrm{a},\mathrm{v}\}$ denote the language, acoustic, and visual modalities. For each $m\in\mathcal{M}$, the input sequence is $X_m$ and the label is $y$. A binary mask vector $\mathbf{\delta}\in\{0,1\}^{|\mathcal{M}|}$ indicates modality availability, where $\delta_m=1$ means observed and $\delta_m=0$ means missing. The observed set is defined as $\mathcal{O}=\{m\in\mathcal{M}\mid \delta_m=1\}$, and the missing set is defined as $\mathcal{M}_{\mathrm{miss}}=\{m\in\mathcal{M}\mid \delta_m=0\}$.

% We define the observed set $\mathcal{O}=\{m\in\mathcal{M}\mid \delta_m=1\}$ and the missing set $\mathcal{M}_{\mathrm{miss}}=\{m\in\mathcal{M}\mid \delta_m=0\}$.

% Let $\mathcal{S}=\{\mathrm{l},\mathrm{a},\mathrm{v}\}$ denote the modality set and
% $\mathcal{K}=\{\mathrm{low},\mathrm{mid},\mathrm{high}\}$ denote semantic levels.
% We represent modality availability by a binary vector $\boldsymbol{\delta}\in\{0,1\}^{|\mathcal{S}|}$.

During training, missingness is simulated by masking modalities according to predefined patterns. A non-masked forward pass on the same sample provides full-modality targets for reconstruction supervision. These targets are treated as constants and detached from gradients to avoid trivial coupling between the reconstruction objective and the target features.

\subsection{Hierarchical Semantic Decoupling (HSD)}
Standard MSA encoders typically extract only the final-layer output, which collapses intermediate semantics into a single representation. To expose the depth-wise semantic progression, HSD partitions the encoder of each modality $m \in \{\mathrm{l},\mathrm{a},\mathrm{v}\}$ into three cascaded blocks. Let the backbone encoder for modality $m$ consist of \textcolor{red}{$L_{\mathrm{m}}$} Transformer layers.

In our implementation, the layer boundaries are defined in a modality-wise manner. 
For each modality-specific encoder $E_m$ with depth $L_m$, we set:
\begin{equation}
l_1^m=\left\lfloor \frac{L_m}{3}\right\rfloor,\quad
l_2^m=\left\lfloor \frac{2L_m}{3}\right\rfloor,\quad m\in\{l,a,v\}.
\end{equation}
Since the three modality-specific encoders have the same depth in our implementation, i.e., $L_l=L_a=L_v=12$, the resulting boundaries are numerically identical, with layers 1--4, 5--8, and 9--12 corresponding to the low-, mid-, and high-level blocks, respectively.

It should be emphasized that this equal-depth split is not based on the assumption that language, acoustic, and visual modalities share identical absolute semantic depth. Instead, it provides a reproducible, modality-neutral, and architecture-search-free way to expose relative abstraction stages within each modality-specific encoder. For language, low-level features mainly encode lexical and local phrasal cues, mid-level features capture compositional and contextual patterns, and high-level features summarize utterance-level affective semantics. For acoustic signals, low-level features describe local physical attributes such as pitch, energy, and spectral patterns, mid-level features capture prosodic and temporal dynamics, 
and high-level features encode vocal affect and speaking style. For visual signals, 
low-level features capture local facial or motion cues, mid-level features represent 
action-unit or gesture-level structures, and high-level features summarize visual 
affect and interaction intent.

Therefore, HSD does not force heterogeneous modalities to have identical semantic meaning at the same layer index. Instead, it preserves a comparable relative hierarchy within each modality, which provides structured streams for cross-level reconstruction and hierarchical fusion. We adopt the equal-depth rule as the default setting to avoid dataset-specific boundary tuning, and examine its sensitivity in Appendix F.

We partition these layers into three continuous stages, namely the low-level block $E_m^{Low}$ from layer $1$ to $l_1^m$, the mid-level block $E_m^{Mid}$ from layer $l_{1}^m+1$ to $l_2^m$, and the high-level block $E_m^{High}$ from layer $l_{2}^m+1$ to $L_{\text{m}}$. The hierarchical feature extraction is:
\begin{equation}
\begin{aligned}
F_m^{Low} &=E_m^{Low}(X_m),F_m^{Mid}=E_m^{Mid}(F_m^{Low}),\\ F_m^{High}=E_m^{High}(F_m^{Mid})
\end{aligned}
\end{equation}
where $X_m$ is the input sequence, $T_m$ is its length, and $d$ is the feature width. $F_m^{Low} \in \mathbb{R}^{T_m \times d}$ captures physical patterns, $F_m^{Mid}$ encodes structural units, and $F_m^{High}$ represents abstract affective semantics. This explicit cascade preserves the semantic hierarchy that will be used for reconstruction and fusion.

\subsection{Cross-Modal Cross-Level Semantic Reconstruction}
Imputing missing modalities at a single feature stage is often ill-posed because low-level cues are highly modality-specific and admit many plausible completions. HSL reconstructs missing modalities in a top-down manner, from high-level semantics to mid-level and low-level details. For each missing-modality $B \in \mathcal{M}_{\mathrm{miss}}$, CM-CLSR outputs reconstructed features $\{\hat{F}_B^{High}, \hat{F}_B^{Mid}, \hat{F}_B^{Low}\}$. The high-level reconstruction $\hat{F}_B^{High}$ is used as a high-level anchor because it provides a semantic constraint that conditions and stabilizes the subsequent reconstruction at lower levels. When multiple modalities are missing, we apply the same procedure independently to each missing-modality, and context is always computed from $\mathcal{O}$.

\subsubsection{High-Level Anchor Generation}
% We first generate the high-level representation $\hat{F}_B^{High}$ for the missing modality $B$. 
The high-level representation $\hat{F}_B^{High}$ is first generated for the missing modality $B$. High-level affective semantics are more alignable across modalities than low-level physical cues; therefore, we construct a cross-modal high-level context and generate $\hat{F}_B^{High}$ as an anchor to constrain subsequent mid/low-level reconstruction. The resulting $\hat{F}_B^{High}$ then serves as the conditioning signal for reconstructing $\hat{F}_B^{Mid}$ and $\hat{F}_B^{Low}$ through semantic injection.

Let $\mathcal{O}$ denote the observed modalities. We concatenate their high-level features to form the semantic context.
\begin{equation}
    C_{\mathcal{O}}^{High}=\mathrm{Concat}\big(\{F_A^{\text{High}}\mid A\in \mathcal{O}\}\big)\in\mathbb{R}^{(\sum_{A\in \mathcal{O}}T_A)\times d}
\end{equation}

The high-level anchor in CM-CLSR is constructed from the currently observed modality set $\mathcal{O}$, rather than from a pre-defined privileged modality. Therefore, HSL does not assume that language, acoustic, or visual modality is always a better anchor. Different modalities provide different types of affective evidence: language usually carries explicit lexical and compositional sentiment cues, acoustic signals provide prosodic and arousal-related cues such as pitch, energy, rhythm, and speaking style, and visual signals provide facial expression, gesture, and interaction-state cues. Thus, the reliability of an anchor is sample-dependent and condition-dependent, depending on the quality, availability, and task relevance of the observed modalities. When only one modality is observed, the reconstruction is necessarily conditioned on this single source. If this source is severely noisy or weakly informative, the generated anchor may contain larger uncertainty, which can affect subsequent mid- and low-level reconstruction. This is an inherent limitation of missing-modality reconstruction rather than a modality-specific assumption of HSL. To mitigate this uncertainty, CM-CLSR starts from high-level task-related semantics, HCES regularizes level-wise semantic consistency, and HIF aggregates observed and reconstructed modality-level tokens without forcing any single modality to dominate.

To generate modality-specific high-level tokens for $B$ while keeping the token layout compatible with its encoder, we introduce learnable queries $Q_B^{High} \in \mathbb{R}^{T_B \times d}$ and compute cross-attention
\begin{equation}
\hat{F}_B^{High} = \mathrm{Softmax}\!\left(\frac{(Q_B^{High} W_Q) (C_{\mathcal{O}}^{High} W_K)^T}{\sqrt{d_k}}\right) (C_{\mathcal{O}}^{High} W_V),
\end{equation}
where $W_Q$, $W_K$, and $W_V$ are learnable projections and $d_k$ is the key dimension. The obtained $\hat{F}_B^{High}$ is the high-level anchor. It captures semantics of modality $B$ under the constraint of observed high-level context, and it will be used as the query signal in the next stages.

\subsubsection{Top-Down Reconstruction via Semantic Injection}
Given the high-level anchor $\hat{F}_B^{High}$, we reconstruct mid-level and low-level features sequentially. For $k \in \{Mid, Low\}$, the observed evidence at level $k$ is defined as:
\begin{equation}
C_{\mathcal{O}}^k = \mathrm{Concat}\big(\{F_A^k \mid A \in \mathcal{O}\}\big).
\end{equation}
$\hat{F}_B^k$ is reconstructed by a Semantic Injection Decoder (SID) that uses the upper-level reconstruction as the query and the same-level observed evidence as key and value. Let $\uparrow(Mid)=High$ and $\uparrow(Low)=Mid$. For $k\in\{Mid,Low\}$, we reconstruct: 
\begin{equation}
\hat{F}_B^{k}= \mathrm{SID}^{k}\!\left(\hat{F}_B^{\uparrow(k)},\, C_O^{k}\right).
\end{equation}
Each $\mathrm{SID}^k$ is implemented by a multi-head cross-attention (MHCA) block followed by feed-forward (FFN) refinement with layer normalization (LN):
\begin{align}
Z &= \mathrm{MHCA}\big(Q=\hat{F}_B^{\uparrow(k)}W_Q^k, \ K=C_{\mathcal{O}}^kW_K^k, \ V=C_{\mathcal{O}}^kW_V^k\big), \\
\tilde{Z} &= \mathrm{LN}\big(\hat{F}_B^{\uparrow(k)} + Z\big), \hat{F}_B^k= \mathrm{LN}\big(\tilde{Z} + \mathrm{FFN}(\tilde{Z})\big).
\end{align}

Using $\hat{F}_B^{\uparrow(k)}$ as the query enforces cross-level semantic consistency because injected details at level $k$ must be compatible with the upper-level anchor. This top-down design reduces the search space of missing-modality completion and mitigates over-smoothed reconstructions under high missingness.

% Let $\uparrow(\mathrm{Mid})=\mathrm{High}$ and 
% $\uparrow(\mathrm{Low})=\mathrm{Mid}$. 
% For $k\in\{\mathrm{Mid},\mathrm{Low}\}$, we reconstruct:
% \begin{equation}
% \hat{F}^{k}_{B}
% =
% \mathrm{SID}_{k}
% \left(
% \hat{F}^{\uparrow(k)}_{B}, C^{k}_{\mathcal{O}}
% \right).
% \end{equation}

% Each $\mathrm{SID}_{k}$ is implemented by a cross-attention block followed by feed-forward refinement:
% \begin{equation}
% Z =
% \mathrm{MHCA}
% \left(
% Q=\hat{F}^{\uparrow(k)}_{B} W^{k}_{Q},
% K=C^{k}_{\mathcal{O}} W^{k}_{K},
% V=C^{k}_{\mathcal{O}} W^{k}_{V}
% \right),
% \end{equation}
% \begin{equation}
% \tilde{Z}
% =
% \mathrm{LN}
% \left(
% \hat{F}^{\uparrow(k)}_{B}+Z
% \right),
% \quad
% \hat{F}^{k}_{B}
% =
% \mathrm{LN}
% \left(
% \tilde{Z}+\mathrm{FFN}(\tilde{Z})
% \right).
% \end{equation}

\subsection{Hierarchical Cascade Error Supervisor (HCES)}
\label{sec:hces}
Top-down reconstruction may accumulate errors across levels. HCES regularizes the reconstruction hierarchy and prevents semantic collapse by enforcing level-wise semantic consistency via prototypical supervision.

For each level $k \in \{Low,Mid,High\}$, we maintain a learnable prototype bank $\mathbf{P}^k = \{p_c^k \in \mathbb{R}^{d}\}_{c=0}^{C-1}$, where $C$ is a hyperparameter controlling the granularity of semantic prototypes. For classification, we set $C$ to the class count unless otherwise stated. For regression tasks, the discretization is used only for the auxiliary prototype supervision in HCES, while the original task loss \(\mathcal{L}_{\mathrm{task}}\) is still optimized on continuous sentiment scores. Therefore, HCES does not quantize the final sentiment prediction or convert regression into a classification problem. Instead, it provides coarse semantic anchors to regularize level-wise observed and reconstructed features. Samples within the same bin are encouraged to share a similar high-level semantic region, but their fine-grained sentiment differences are still modeled by the continuous latent representation, the reconstruction objective, the hierarchical fusion module, and the final regression loss. Since different datasets have different score ranges, we first normalize the sentiment score into \([0,1]\), $\tilde{y}=\frac{y-y_{\min}}{y_{\max}-y_{\min}}$, where \(y_{\min}\) and \(y_{\max}\) denote the minimum and maximum sentiment scores of the corresponding dataset. The prototype index is then assigned by $c(y)=\min\left(\left\lfloor C\tilde{y}\right\rfloor, C-1\right)$. Thus, the same prototype count \(C\) does not impose identical raw-score intervals across datasets with different label ranges.

We adopt uniform ordinal binning because it preserves the metric structure of the sentiment scale and keeps the semantic meaning of each prototype interval comparable across datasets. Density-aware binning may balance the number of samples assigned to each prototype and stabilize prototype updates when the label distribution is highly skewed. However, it also produces variable-width intervals, which may merge sparse but semantically important affective extremes into overly broad bins. We therefore use uniform binning as the default choice and regard density-aware prototype construction as a promising future extension.

% For regression, we discretize labels into $C$ uniform bins. 

At each level $k$, we pool both observed and reconstructed features
\begin{equation}
\mathcal{F}^k = \{f_m^k \mid m \in \mathcal{O}\} \cup \{\hat{f}_m^k \mid m \in \mathcal{M}_{\mathrm{miss}}\},
\end{equation}
where $f_m^k=\mathrm{Pool}(F_m^k)$ and $\hat{f}_m^k=\mathrm{Pool}(\hat{F}_m^k)$, \textcolor{red}{and Pool(·) denotes average pooling over the temporal dimension}. We optionally apply $\ell_2$ normalization to both features and prototypes before similarity computation for numerical stability. For each $z \in \mathcal{F}^k$, we define the prototype-based loss
\begin{equation}
\ell_{\mathrm{proto}}(z, y, k) =
- \log \frac{\exp\big(z^\top p_{c(y)}^k / \tau\big)}
{\sum_{c=0}^{C-1}\exp\big(z^\top p_c^k / \tau\big)}.
\end{equation}
The temperature $\tau$ controls the sharpness of the prototype distribution. Smaller $\tau$ yields more selective assignments and stronger inter-prototype separation, which improves resistance to semantic averaging, while larger $\tau$ provides softer supervision.

The final HCES objective aggregates all levels
\begin{equation}
\mathcal{L}_{\mathrm{HCES}} =
\sum_{k \in \{Low,Mid,High\}} \sum_{z \in \mathcal{F}^k} \ell_{\mathrm{proto}}(z, y, k).
\end{equation}
By constraining $\hat{F}_B^{High}$ before it conditions lower-level reconstruction and by enforcing consistency at all levels, HCES reduces error propagation in the reconstruction cascade.

\subsection{Hierarchical Interactive Fusion (HIF)}
After reconstruction, we obtain hierarchical features for all modalities at three semantic levels. HIF jointly models cross-modal and cross-level dependencies and produces a task-ready fused representation.
\label{sec:hif}

\paragraph{Composite interaction sequence}
We flatten level-wise features into a unified token sequence. For each modality $m \in \mathcal{M}$ and level $k \in \{Low,Mid,High\}$, we denote the available feature stream by $\tilde{F}_m^k$, where $\tilde{F}_m^k = F_m^k$ if $m \in \mathcal{O}$ and $\tilde{F}_m^k = \hat{F}_m^k$ if $m \in \mathcal{M}_{\mathrm{miss}}$.
We introduce learnable modality embeddings $e_{\mathrm{mod}}^m \in \mathbb{R}^{d}$ and level embeddings $e_{\mathrm{level}}^k \in \mathbb{R}^{d}$ to encode token origin and semantic scale, which helps the fusion module distinguish modality and level identities and reduces semantic mixing.
A learnable aggregation token $x_{\mathrm{cls}} \in \mathbb{R}^{1 \times d}$ is prepended to collect global evidence. The input sequence is
% \begin{equation}
% \mathcal{S}_{\mathrm{in}} =
% \Big[
% x_{\mathrm{cls}};
% \ \tilde{F}_m^k + e_{\mathrm{mod}}^m + e_{\mathrm{level}}^k
% \Big]_{m \in \mathcal{M},\ k \in \{Low,Mid,High\}} \in \mathbb{R}^{N \times d},
% \end{equation}
\begin{equation}
\begin{aligned}
\mathcal{S}_{\mathrm{in}}
&=
\Big[
x_{\mathrm{cls}};\,
\big\{
\widetilde{F}_m^k
+ e_{\mathrm{mod}}^m
+ e_{\mathrm{level}}^k
\big\}_{\substack{
m\in\mathcal{M},\\
k\in\{\mathrm{Low},\mathrm{Mid},\mathrm{High}\}
}}
\Big],\\
&\in \mathbb{R}^{N\times d},
\end{aligned}
\label{eq:input_sequence}
\end{equation}
where $N$ is the total number of tokens after concatenation.

\paragraph{Affinity matrix in fusion}
We feed $\mathcal{S}_{\mathrm{in}}$ into a Transformer encoder $\mathcal{T}_{\mathrm{fuse}}$ with $L_{\mathrm{fuse}}$ layers. Because tokens are indexed by both modality and level, self-attention naturally captures cross-modal interactions, cross-level interactions, and their mixed dependencies within the same routing mechanism.
At layer $t$, given $\mathcal{S}^{t-1} \in \mathbb{R}^{N \times d}$, we compute
\begin{equation}
Q^t = \mathcal{S}^{t-1} W_Q^t,\quad
K^t = \mathcal{S}^{t-1} W_K^t,\quad
V^t = \mathcal{S}^{t-1} W_V^t,
\end{equation}
and define the affinity matrix
\begin{equation}
\mathcal{A}^t =
\mathrm{Softmax}\!\left(\frac{Q^t (K^t)^{\top}}{\sqrt{d_k}}\right)\in\mathbb{R}^{N\times N}.
\end{equation}
$\mathcal{A}^t$ parameterizes a routing distribution over all token pairs. Each entry $\mathcal{A}^t_{ij}$ measures how much information token $i$ should aggregate from token $j$. The attention output is computed by value aggregation $\mathrm{Attn}^t = \mathcal{A}^t V^t \in \mathbb{R}^{N \times d}.$ We implement multi-head attention and apply residual connections and feed-forward refinement
\begin{align}
\tilde{\mathcal{S}}^{t} &= \mathrm{LN}\!\big(\mathcal{S}^{t-1} + \mathrm{Attn}^t\big),
\mathcal{S}^{t} = \mathrm{LN}\!\big(\tilde{\mathcal{S}}^{t} + \mathrm{FFN}(\tilde{\mathcal{S}}^{t})\big).
\end{align}
After $L_{\mathrm{fuse}}$ layers, we obtain $\mathcal{S}_{\mathrm{out}}=\mathcal{S}^{L_{\mathrm{fuse}}}$.

\paragraph{Task-ready fused representation}
We use the output embedding of the aggregation token as the global fused representation
\begin{equation}
h = \mathcal{S}_{\mathrm{out}}[0]\in\mathbb{R}^{d}, \\ \hat{y} = \phi(h),
\end{equation}
which summarizes evidence routed from all modality-level tokens via the learned affinity matrices across layers. The prediction is produced by a task-specific head $\phi(\cdot)$, implemented as a linear regressor for regression tasks or an MLP classifier with softmax for classification tasks.

\subsection{Optimization Objective}
\label{sec:objective}
% We train the framework end-to-end with a compound objective. We define a reconstruction loss that supervises generated features against the full-modality targets $F_{m,\mathrm{GT}}^k$ extracted by a non-masked pass and detached from gradients
% \begin{equation}
% \mathcal{L}_{\mathrm{rec}} = \sum_{m \in \mathcal{M}_{\mathrm{miss}}} \sum_{k \in \{Low,Mid,High\}} \| \hat{F}_m^k - F_{m,\mathrm{GT}}^k \|_2^2.
% \end{equation}
To supervise completion under both modality-level absence and token/frame-level erasure, we define a mask-aware reconstruction loss.
For each sample, we obtain the full-modality targets $\{F^{k}_{m,\mathrm{GT}}\}$ by a non-masked forward pass and detach them from gradients to avoid trivial coupling. Let $\mathcal{K}=\{Low,Mid,High\}$ denote the set of semantic levels. We then compute reconstruction loss only on erased positions:
\begin{equation}
\mathcal{L}_{\mathrm{rec}}
=
\sum_{m\in\mathcal{M}}
\sum_{k\in\mathcal{K}}
\left\|
(1-\mathbf{s}_m)\odot\left(\hat F_m^{k}-F^{k}_{m,\mathrm{GT}}\right)
\right\|_F^2,
\label{eq:mask_aware_lrec}
\end{equation}
where $s_m\in\{0,1\}^{T_m}$ is a position mask (1=kept, 0=erased), broadcast along the feature dimension; for modality-level missingness, we set $s_m=0$ for $m\in\mathcal{M}_{\mathrm{miss}}$.

The total objective balances task prediction, reconstruction fidelity, and semantic consistency:
\begin{equation}
\mathcal{L}_{\mathrm{total}} = \mathcal{L}_{\mathrm{task}} + \lambda_1 \mathcal{L}_{\mathrm{rec}} + \lambda_2 \mathcal{L}_{\mathrm{HCES}},
\end{equation}
where $\mathcal{L}_{\mathrm{task}}$ is the task-specific loss, and $\lambda_1$ and $\lambda_2$ balance the auxiliary objectives.

% \begin{table}[t]
% \centering
% \caption{THE STATISTICAL DATA OF THE DATASETS}
% \label{tab:dataset_stats}
% \begin{tabular}{c|c c c c}
% \midrule
% \textbf{Benchmark} & \textbf{Train} & \textbf{Val} & \textbf{Test} & \textbf{Total} \\
% \hline
% \hline
% CMU-MOSI~\cite{Cmu-mosi}       & 1,284  & 229  & 686  & 2,199  \\
% CMU-MOSEI~\cite{Cmu-mosei}      & 16,326 & 1,871 & 4,659 & 22,856 \\
% IEMOCAP-FOUR~\cite{iemocap}   & 4,425  & 553  & 553  & 5,531  \\
% IEMOCAP-SIX~\cite{iemocap}    & 5,945  & 744  & 744  & 7,433  \\
% CH-SIMS~\cite{sims}           & 1,368 & 456 & 457 & 2,281 \\
% \midrule
% \end{tabular}
% \end{table}

\section{Experiments}

\subsection{Benchmarks}
HSL is evaluated on four widely used benchmarks: CMU-MOSI~\cite{Cmu-mosi} and CMU-MOSEI~\cite{Cmu-mosei}, consisting of English YouTube clips with sentiment scores in $[-3, 3]$; CH-SIMS, a Chinese dataset from film and TV with scores in $[-1, 1]$~\cite{sims}, and IEMOCAP~\cite{iemocap}, an acted-dialogue corpus for 4-class and 6-class emotion recognition. We use the official train/validation/test splits of CMU-MOSI, CMU-MOSEI, and CH-SIMS. Since IEMOCAP does not provide a fixed official split in our experimental protocol, we randomly partition its utterances into training, validation, and test sets at an 8:1:1 ratio, following the adopted utterance-level protocol. Details are given in Appendix A.

\subsection{Experimental Settings}

Robustness is evaluated under two complementary missing-modality protocols.

\subsubsection{Intra-modal missingness}

We simulate intra-modal missingness by random erasure within each modality sequence. For a target missing rate $r$, we independently sample a binary mask over positions for each modality and erase a fraction $r$ of positions for that modality in each sample. This sampling is performed separately for language tokens and for acoustic and visual frames, and it preserves the temporal ordering of the remaining positions.
Following prior settings~\cite{P-RMF}, erased acoustic and visual positions are replaced by zero padding, and erased language tokens are replaced by the unknown token \texttt{[UNK]}.
We evaluate missing rates $r \in \{0.0, 0.1, \ldots, 0.9\}$.

\subsubsection{Fixed-modality missingness}
We further evaluate robustness under fixed-modality missingness, where one modality or two modalities are entirely absent at inference time.

\subsubsection{Training protocol and fairness}
We adopt protocol-specific training and evaluation to follow prior protocol comparisons under both missing-modality \textcolor{red}{settings}. These regimes reflect different assumptions.
Under intra-modal missingness, for methods that support missing inputs during training (including ours), we apply the same missingness process to both training and test data for each missing rate $r$; for methods that require complete modalities during training, they are trained on fully observed data and intra-modal missingness is imposed only at test time, following the common protocol in prior work~\cite{P-RMF}. We summarize robustness by reporting the mean performance averaged over all evaluated missing rates. \textcolor{red}{Under fixed-modality missingness, complete-input methods (e.g., Self-MM, DCCA, DCCAE, and DMD) are trained on the fully observed training split, and at test time missing acoustic and visual streams are replaced by zero padding and missing language tokens by [UNK]. Missing-modality methods (e.g., MMIN, GCNet, IMDer, CorrKD, P-RMF, and ROSA) follow the training strategies of their original papers. HSL is trained with the modality-level masking described in Section~\ref{sec:III-A} and evaluated by removing the corresponding modality streams at test time.} We report results for each fixed condition and also the average across all tested conditions. To explicitly quantify the effect of this training-protocol asymmetry, we additionally report an HSL-CompleteTrain control study in Appendix H, where HSL is trained only on complete modalities and evaluated under test-time missingness.

\subsubsection{Evaluation metrics}
Under intra-modal missingness, we report Acc-2, Acc-5, Acc-7, F1, MAE, and Corr on CMU-MOSI and CMU-MOSEI, and Acc-2, Acc-3, Acc-5, F1, MAE, and Corr on CH-SIMS~\cite{sims}. \textcolor{red}{On CMU-MOSI and CMU-MOSEI, Acc-2 and F1 are reported under negative vs. non-negative (Has0) and negative vs. positive (Non0) settings. All HSL F1 scores are computed as Weighted F1, with class weights determined by the ground-truth class supports. Following LNLN~\cite{LNLN}, the best checkpoint is selected separately for each evaluation metric and evaluated at every missing rate. Acc-2 and its corresponding F1 are computed from the same checkpoint, whereas Acc-5, Acc-7, MAE, and Corr may use their respective best checkpoints. Under fixed-modality missingness, we report Non0 F1 on CMU-MOSI and CMU-MOSEI, following P-RMF~\cite{P-RMF}, and support-weighted F1 on IEMOCAP for both the 4-class and 6-class settings.}

\subsubsection{Implementation details}
All experiments are conducted with PyTorch on a single A100 GPU (80GB). HSL results are averaged over five random seeds; baseline results come from the cited papers or from results reported in prior studies (e.g.,~\cite{LNLN,P-RMF}) unless otherwise specified. The detailed setup and parameters, as well as the baseline details, are provided in Appendices B and C, respectively.

\begin{table*}[ht]
\centering
\caption{Robustness comparison on MOSI and MOSEI under intra-modal missingness. Results are averaged over $r \in \{0.0,0.1,\ldots,0.9\}$.}
\label{tab:CMU-Intra-Modality_results}
\setlength{\tabcolsep}{4.5pt}
\renewcommand{\arraystretch}{1.1}
\begin{tabular}{lcccccc|cccccc}
\toprule
& \multicolumn{6}{c|}{\textbf{CMU-MOSI}} & \multicolumn{6}{c}{\textbf{CMU-MOSEI}} \\
\cmidrule(lr){2-7} \cmidrule(lr){8-13}
\textbf{Model} & Acc-2 & F1 & Acc-5 & Acc-7 & MAE ($\downarrow$) & Corr 
               & Acc-2 & F1 & Acc-5 & Acc-7 & MAE ($\downarrow$) & Corr \\
\midrule
MISA~\cite{misa}     & 70.33/71.49 & 70.00/71.28 & 33.08 & 29.85 & 1.085 & 0.524
         & 75.82/71.27 & 68.73/63.85 & 39.39 & \textcolor{red}{38.92} & 0.780 & 0.503 \\
Self-MM~\cite{self-mm}  & 69.26/70.51 & 67.54/66.60 & 34.67 & 29.55 & 1.070 & 0.512
         & 77.42/73.89 & 72.31/68.92 & 45.38 & 44.70 & 0.695 & 0.498 \\
MMIM~\cite{MMIM}     & 67.06/69.14 & 64.04/66.65 & 33.77 & 31.30 & 1.077 & 0.507
         & 75.89/73.32 & 70.32/68.72 & 41.74 & 40.75 & 0.739 & 0.489 \\
CENET~\cite{CENET}    & 67.73/71.46 & 64.85/68.41 & \textcolor{red}{33.62} & 30.38 & 1.080 & 0.504
         & 77.34/74.67 & 74.08/70.68 & 47.83 & 47.18 & 0.685 & 0.535 \\
% TFR-Net~\cite{TFR-Net}  & 66.35/68.15 & 60.06/61.73 & 34.67 & 29.54 & 1.200 & 0.459
%          & 77.23/73.62 & 71.99/68.80 & 34.67 & 46.83 & 0.697 & 0.489 \\
ALMT~\cite{ALMT}     & 68.39/70.40 & 71.80/72.57 & 33.42 & 30.30 & 1.083 & 0.498
         & 77.54/76.64 & 78.03/77.14 & 41.64 & 40.92 & 0.674 & 0.481 \\
LNLN~\cite{LNLN}     & 70.94/72.55 & 71.25/72.73 & 38.27 & 34.26 & 1.046 & 0.527
         & 78.19/76.30 & 79.95/77.77 & 46.17 & 45.42 & 0.692 & 0.530 \\
NIAT~\cite{NIAT}     & 68.02 & 66.13 & - & - & 1.131 & -
         & 77.81 & 75.24 & - & - & 0.69 & - \\
P-RMF~\cite{P-RMF} 
         & 71.53/72.81 & 71.69/72.93 & 38.50 & 34.19 & 1.038 & 0.525
         & 78.83/78.14 & \textbf{80.39}/79.33 & 45.87 & 44.63 & 0.658 & 0.589 \\
DAR~\cite{DAR} 
         & 71.60/73.18 & 71.51/73.15 & 38.65 & 34.47 & 1.069 & 0.520
         & 77.48/78.14 & 77.44/77.51 & 48.01 & 47.02 & 0.666 & 0.583 \\
MIG-HCL~\cite{MIG-HCL} 
         & 72.62 & 72.17 & - & 34.4 & 0.993 & 0.508 
         & 77.73 & 76.72 & - & 48.32 & 0.646 & 0.557 \\
\midrule
\textbf{HSL (Ours)} & \textbf{73.25/\textcolor{red}{74.21}} & \textbf{\textcolor{red}{73.28/73.77}} & \textbf{\textcolor{red}{40.10}} & \textbf{\textcolor{red}{36.76}} & \textbf{0.965} & \textbf{\textcolor{red}{0.563}}
         & \textbf{80.12/81.35} & \textcolor{red}{80.06}/\textbf{81.45} & \textbf{49.76} & \textbf{\textcolor{red}{49.62}} & \textbf{0.635} & \textbf{0.612} \\
\bottomrule
\end{tabular}
\vspace{1mm}

\footnotesize{$1\dagger$ $a/b$ represents binary accuracy and F1 under negative vs. non-negative (Has0) and negative vs. positive (Non0) settings.}\\
\footnotesize{$2\dagger$ Results of HSL are averaged over five random seeds; the standard deviations over seeds on CMU-MOSI are reported in Fig.~\ref{fig:Stability} (Section~\ref{sec:Statistical Stability Analysis})}.\\
\footnotesize{$3\dagger$ Following~\cite{LNLN}, HSL checkpoints are selected according to the evaluation metric. For each binary setting, Acc-2 and its corresponding F1 are computed from the same predictions, whereas other metrics use their respective best checkpoints. HSL's F1 is weighted. Baselines are from the cited papers or~\cite{LNLN,P-RMF}; MISA's CMU-MOSEI Acc-7 (38.92\%) is from our reproduction, and CENET's CMU-MOSI Acc-5 (33.62\%) follows the corrigendum of~\cite{LNLN}}.

\end{table*}

\begin{table}[ht]
\centering
\caption{Robustness comparison on CH-SIMS under intra-modal missingness. Results are averaged over $r \in \{0.0,0.1,\ldots,0.9\}$.}
\label{tab:sims_intra_robustness}
\setlength{\tabcolsep}{4pt}
\renewcommand{\arraystretch}{1.1}
\begin{tabular}{lcccccc}
\toprule
\textbf{Model} & \textbf{Acc-2} & \textbf{F1} & \textbf{Acc-3} & \textbf{Acc-5} & \textbf{MAE} ($\downarrow$) & \textbf{Corr} \\
\midrule
MISA~\cite{misa}    & 72.71 & 66.30 & 56.87 & 31.53 & 0.539 & 0.348 \\
Self-MM~\cite{self-mm}             & 72.81 & 68.43 & 56.75 & 32.28 & 0.508 & 0.376 \\
MMIM~\cite{MMIM}                & 69.86 & 66.21 & 52.76 & 31.81 & 0.544 & 0.339 \\
CENET~\cite{CENET}               & 68.13 & 57.90 & 53.17 & 22.29 & 0.589 & 0.107 \\
TFR-Net~\cite{TFR-Net}             & 68.13 & 58.70 & 52.89 & 26.52 & 0.661 & 0.169 \\
ALMT~\cite{ALMT}                & 71.85 & \textcolor{red}{67.49} & 56.47 & 34.16 & 0.509 & 0.372 \\
LNLN~\cite{LNLN}    & \textcolor{red}{73.15} & \textcolor{red}{69.18} & \textcolor{red}{57.32} & \textcolor{red}{34.68} & \textcolor{red}{0.515} & \textcolor{red}{0.387} \\
P-RMF~\cite{P-RMF}   & 73.64 & 74.65 & 54.75 & 34.83 & 0.500 & 0.414 \\
\midrule
\textbf{HSL (Ours)} & \textbf{75.25} & \textbf{\textcolor{red}{75.26}} & \textbf{58.92} & \textbf{37.15} & \textbf{0.476} & \textbf{0.452} \\
\bottomrule

\end{tabular}
\vspace{1mm}

\footnotesize{$\dagger$ Following~\cite{LNLN}, HSL checkpoints are selected according to the evaluation metric. HSL's F1 is weighted. Baseline results are taken from the cited papers or the results reported in~\cite{LNLN,P-RMF}, with the LNLN row and the F1 of ALMT updated according to the LNLN authors' corrigendum.}
\end{table}

% Detailed parameters are provided in Table~\ref{tab:hyperparameters}.
% the text dimension $d_t$ is set to 768, 
% We use Adam with a learning rate of $1\times 10^{-4}$, a batch size of 32, and 100 training epochs.

% \subsection{Model Zoo}
% To evaluate HSL under missing-modality conditions, we compare it with representative baselines spanning three categories.
% \textbf{General fusion frameworks} assume complete modalities and provide references for feature extraction and fusion. They include DCCAE~\cite{DCCAE}, MISA~\cite{misa}, Self-MM~\cite{self-mm}, MMIM~\cite{MMIM}, CENET~\cite{CENET}, ALMT~\cite{ALMT}, DMD~\cite{dmd}, and related Transformer-style fusion designs.
% \textbf{Translation and reconstruction methods} explicitly generate or recover missing signals. Translation-based models include MCTN, TransM, MMIN~\cite{MMIN}, and CIF-MMIN. Reconstruction-based models include IMDer~\cite{IMDer}, TFR-Net~\cite{TFR-Net}, GCNet~\cite{GCNet}, and DAR~\cite{DAR}.
% \textbf{Robust learning frameworks} improve robustness through uncertainty modeling, noise adaptation, or knowledge transfer. They include LNLN~\cite{LNLN}, NIAT~\cite{NIAT}, SMIL, EMT~\cite{EMT}, CorrKD~\cite{CorrKD}, GGMD~\cite{GGMD}, P-RMF~\cite{P-RMF}, and ROSA~\cite{ROSA}.

\begin{table}[t] 
\centering 
\caption{RESULTS ON THE MOSI, MOSEI, and IEMOCAP-FOUR/-SIX WITH FIXED MISSING MODALITIES. \textcolor{red}{On MOSI and MOSEI, F1 is the Non0 F1, as in P-RMF~\cite{P-RMF}; on IEMOCAP, it is the weighted F1.}} \label{tab:fixed_missing_results}
\small
\scriptsize
\setlength{\tabcolsep}{3pt}
\renewcommand{\arraystretch}{1.1}

\begin{adjustbox}{max width=\columnwidth}
\begin{tabular}{M{1.7cm}|cccccccc}
\midrule
\multirow{2}{*}{\centering\textbf{Models}} & \multicolumn{8}{c}{\textbf{Test Conditions (F1-score, \%)}} \\
\cline{2-9}
& \rotatebox{90}{$\{l, v, a \}$}
& \rotatebox{90}{$\{l, v\}$}
& \rotatebox{90}{$\{l, a\}$}
& \rotatebox{90}{$\{v, a\}$}
& \rotatebox{90}{$\{l\}$}
& \rotatebox{90}{$\{v\}$}
& \rotatebox{90}{$\{a\}$}
& \rotatebox{90}{\textbf{Average}} \\

\midrule\midrule
\multicolumn{9}{c}{\textbf{CMU-MOSI}} \\
% \midrule
Self-MM~\cite{self-mm} & 84.64 & 74.97 & 69.81 & 47.12 & 67.80 & 38.52 & 40.95 & 60.54 \\
DCCAE~\cite{DCCAE} & 66.76 & 65.43 & 62.25 & 42.39 & 61.78 & 41.45 & 41.30 & 54.48 \\
DMD~\cite{dmd} & 84.50 & 68.45 & 70.51 & 50.47 & 68.97 & 42.26 & 43.33 & \textcolor{red}{61.21} \\
MMIN~\cite{MMIN} & 85.20 & 84.76 & 83.50 & 45.51 & 82.39 & 43.86 & 44.25 & 67.07 \\
NIAT~\cite{NIAT} & 83.16 & 81.75 & 82.03 & 43.58 & 80.72 & 42.79 & 42.10 & 65.16 \\
GCNet~\cite{GCNet} & 83.20 & 83.58 & 84.73 & 70.02 & \textcolor{red}{80.91} & 59.67 & \textcolor{red}{65.07} & 75.31 \\
CIF-MMIN~\cite{CIF-MMIN} & 84.27 & 82.48 & 81.62 & 54.70 & 81.16 & 51.09 & 52.02 & 69.62 \\
IMDer~\cite{IMDer} & 84.71 & 81.67 & 82.53 & 57.30 & 81.40 & 56.71 & 55.29 & 71.37 \\
EMT~\cite{EMT} & 82.43 & 83.30 & 82.64 & 55.90 & 80.37 & 53.63 & 51.50 & 69.97 \\
GGMD~\cite{GGMD} & 85.70 & 84.39 & 83.53 & 54.60 & 81.74 & 52.88 & 51.24 & 70.58 \\
CorrKD~\cite{CorrKD} & 83.94 & 82.41 & \textcolor{red}{83.56} & 73.74 & 81.20 & 60.72 & 66.52 & 76.01 \\
P-RMF~\cite{P-RMF} & 84.37 & 81.94 & 82.10 & 73.11 & 81.36 & \textbf{70.32} & 71.44 & 77.81 \\
ROSA~\cite{ROSA} & 86.30 & 85.39 & 85.13 & 83.81 & 83.64 & 56.79 & 81.91 & 80.42 \\
\textbf{HSL(Ours)} & \textbf{86.85} & \textbf{86.12} & \textbf{85.94} & \textbf{84.56} & \textbf{84.21} & 59.35 & \textbf{82.48} & \textbf{81.36} \\

\midrule\midrule
\multicolumn{9}{c}{\textbf{CMU-MOSEI}} \\
% \midrule
Self-MM~\cite{self-mm} & 83.69 & 74.62 & 75.91 & 49.52 & 71.53 & 37.61 & 43.57 & 62.35 \\
DCCAE~\cite{DCCAE} & 75.70 & 73.16 & 73.82 & 45.03 & 71.19 & 43.84 & 41.47 & 60.60 \\
DMD~\cite{dmd} & 84.78 & 72.45 & 74.78 & 52.70 & 70.26 & 39.84 & 46.18 & 63.00 \\
MMIN~\cite{MMIN} & 85.78 & 83.55 & 82.93 & 49.35 & 82.21 & 47.63 & 46.08 & 68.22 \\
NIAT~\cite{NIAT} & 86.35 & 85.31 & 85.10 & 48.57 & 83.29 & 45.76 & 44.43 & 68.40 \\
GCNet~\cite{GCNet} & 82.35 & 81.15 & 81.96 & 69.21 & 80.52 & 61.83 & 66.54 & 74.79 \\
CIF-MMIN~\cite{CIF-MMIN} & 85.33 & 83.36 & 82.70 & 53.69 & 81.94 & 50.70 & 51.38 & 69.87 \\
IMDer~\cite{IMDer} & 84.71 & 82.39 & 83.41 & 53.30 & 82.77 & 51.83 & 50.47 & 69.84 \\
EMT~\cite{EMT} & 85.17 & 84.45 & 83.81 & 52.39 & 83.04 & 50.79 & 51.28 & 70.13 \\
GGMD~\cite{GGMD} & 87.07 & 85.52 & 83.10 & 52.67 & 81.45 & 49.96 & 50.39 & 70.02 \\
CorrKD~\cite{CorrKD} & 82.16 & 81.28 & 81.74 & 71.92 & 80.76 & 62.30 & 66.09 & 75.18 \\
P-RMF~\cite{P-RMF} & 85.48 & 85.17 & 84.61 & 76.88 & 81.91 & \textbf{73.19} & 75.91 & 80.45 \\
ROSA~\cite{ROSA} & 89.56 & 87.32 & 86.64 & 84.19 & 84.42 & 54.38 & 79.04 & 80.79 \\
\textbf{HSL(Ours)} & \textbf{89.89} & \textbf{87.95} & \textbf{87.41} & \textbf{85.03} & \textbf{84.97} & 56.82 & \textbf{79.65} & \textbf{81.67} \\

\midrule\midrule
\multicolumn{9}{c}{\textbf{IEMOCAP-FOUR}} \\
% \midrule
DCCA~\cite{DCCA} & 62.27 & 61.73 & 60.05 & 42.78 & 59.30 & 44.73 & 42.91 & 53.40 \\
DCCAE~\cite{DCCAE} & 63.12 & 62.55 & 60.13 & 42.90 & 60.03 & 45.48 & 45.56 & 54.25 \\
MMIN~\cite{MMIN} & 76.35 & 75.62 & 75.89 & 51.72 & 73.45 & 47.70 & 45.31 & 63.72 \\
NIAT~\cite{NIAT} & 76.87 & 74.35 & 74.79 & 50.03 & 73.32 & 48.55 & 47.79 & 63.67 \\
CIF-MMIN~\cite{CIF-MMIN} & 74.45 & 73.15 & 72.69 & 47.52 & 72.25 & 45.88 & 46.28 & 61.75 \\
IMDer~\cite{IMDer} & 77.65 & 76.40 & 75.64 & 51.78 & 73.26 & 47.89 & 47.05 & 64.24 \\
EMT~\cite{EMT} & 76.34 & 75.19 & 74.87 & 50.36 & 72.53 & 48.10 & 49.13 & 63.79 \\
GGMD~\cite{GGMD} & 78.41 & 77.53 & 76.06 & 52.78 & 75.80 & 51.38 & 49.75 & 65.96 \\
ROSA~\cite{ROSA} & 81.93 & 79.85 & 78.63 & 73.96 & 75.90 & 52.17 & 69.30 & 73.11 \\
\textbf{HSL(Ours)} & \textbf{82.68} & \textbf{80.76} & \textbf{79.45} & \textbf{75.82} & \textbf{76.43} & \textbf{54.89} & \textbf{70.51} & \textbf{74.36} \\

\midrule\midrule
\multicolumn{9}{c}{\textbf{IEMOCAP-SIX}} \\
% \midrule
DCCA~\cite{DCCA} & 42.63 & 41.79 & 40.34 & 33.54 & 39.93 & 31.80 & 29.37 & 37.06 \\
DCCAE~\cite{DCCAE} & 45.53 & 44.76 & 42.20 & 34.84 & 41.17 & 31.97 & 30.49 & 38.71 \\
MMIN~\cite{MMIN} & 57.02 & 56.61 & 55.74 & 36.85 & 54.30 & 34.92 & 33.87 & 47.04 \\
NIAT~\cite{NIAT} & 56.39 & 54.74 & 55.16 & 34.64 & 54.28 & 33.40 & 32.69 & 45.90 \\
CIF-MMIN~\cite{CIF-MMIN} & 57.34 & 56.49 & 55.34 & 37.72 & 54.07 & 36.64 & 35.71 & 47.62 \\
IMDer~\cite{IMDer} & 57.49 & 55.34 & 56.90 & 37.23 & 54.03 & 35.58 & 34.21 & 47.25 \\
EMT~\cite{EMT} & 56.17 & 56.01 & 54.79 & 39.71 & 53.62 & 36.05 & 33.47 & 47.12 \\
GGMD~\cite{GGMD} & 58.19 & 57.48 & 55.35 & 38.40 & 52.25 & 36.03 & 35.83 & 47.65 \\
ROSA~\cite{ROSA} & 62.65 & 58.38 & 58.17 & 55.76 & 56.33 & 36.25 & 55.39 & \textcolor{red}{54.70} \\
\textbf{HSL(Ours)} & \textbf{63.41} & \textbf{59.76} & \textbf{59.22} & \textbf{57.04} & \textbf{57.15} & \textbf{38.92} & \textbf{56.83} & \textbf{56.05} \\

\midrule
% \multicolumn{9}{l}{\footnotesize The best performance is highlighted in bold.}
\end{tabular}
\end{adjustbox}
\end{table}

\subsection{Robustness Comparison on Intra-Modal Missingness}
\label{sec:exp_intra_robustness}
We evaluate robustness against random information erasure within each modality on CMU-MOSI, CMU-MOSEI, and CH-SIMS.
As shown in Tables~\ref{tab:CMU-Intra-Modality_results} and Table~\ref{tab:sims_intra_robustness}, HSL achieves consistently strong performance across all metrics.
\textcolor{red}{In particular, HSL attains the highest Pearson correlation between the predicted and ground-truth sentiment scores (Corr of 0.563 on CMU-MOSI and 0.612 on CMU-MOSEI). On CH-SIMS, HSL achieves the best results among the compared methods across all six evaluation metrics. Compared with P-RMF, which obtains the highest F1 among the listed baselines, HSL improves Acc-2, F1, and Acc-5 by 1.61, 0.61, and 2.32 percentage points, respectively. Compared with LNLN using the authors' corrected result, HSL achieves gains of 2.10 percentage points in Acc-2 and 6.08 percentage points in F1.} More detailed performance curves and experimental results under different missing rates are provided in Appendix D.

Under intra-modal missingness, random erasure disrupts local temporal continuity and aggravates cross-modal asynchrony, which degrades direct fusion of corrupted streams. HSL mitigates this by completing hierarchical features before fusion: high-level anchors constrain top-down completion of mid/low-level details, while HCES enforces level-wise semantic consistency to prevent drift. Consequently, HIF operates on completed and semantically regularized tokens, improving robustness across missing rates.

\subsection{Robustness Comparison on Fixed-Modality Missingness}
\label{sec:exp_fixed_robustness}
We further evaluate robustness when one modality or two modalities are entirely absent at inference time. Table~\ref{tab:fixed_missing_results} demonstrates that HSL achieves competitive performance across all four settings, covering coarse-grained sentiment analysis (MOSI and MOSEI) and fine-grained emotion recognition (IEMOCAP-FOUR and IEMOCAP-SIX). HSL also remains stable in the text-absent conditions $\{\mathrm{v},\mathrm{a}\}$ and $\{\mathrm{a}\}$, which is a challenging case because language often provides dominant semantic supervision.

Fixed-modality missingness differs from intra-modal missingness because an entire semantic source becomes unobservable, which makes simple down-weighting strategies insufficient.
Uncertainty-aware methods, such as ROSA, focus on estimating modality reliability, but they cannot recover missing semantic content once a primary modality is removed.
HSL uses active semantic synthesis to reduce representation gaps created by missing modalities.
HSD preserves the hierarchical integrity of each modality so that available modalities contribute at multiple semantic scales.
CM-CLSR reconstructs the missing modality hierarchy using high-level anchors from observed modalities, which helps approximate missing categorical and affective cues.
HCES enforces level-wise semantic consistency during reconstruction, which pulls reconstructed features toward their correct semantic prototypes at each level.
Consequently, HIF can aggregate complementary evidence from all available and reconstructed tokens, improving robustness under highly incomplete inputs.

\textcolor{red}{HSL uses protocol-specific training: modality-level masking for fixed-modality missingness and token-/frame-level erasure for intra-modal missingness (Sections III-A and IV-B3). Thus, the ({l,v,a}) result differs from (r=0) in Table~\ref{tab:CMU-Intra-Modality_results} and Table~\ref{tab:sims_intra_robustness}.}

\subsection{Ablation Studies}
\label{sec:ablation}
To verify the effectiveness of HSL components, reconstruction strategies, fusion mechanisms, and training objectives, we conduct ablation studies on CMU-MOSI, CMU-MOSEI, and CH-SIMS under intra-modal missingness with $r \in \{0.0, 0.1,\ldots,0.9\}$.
We report the average F1 (Non0 on CMU-MOSI and CMU-MOSEI) and MAE in Table~\ref{tab:ablation_hsl_missingness}.

\begin{table}[t]
\caption{Comprehensive ablation studies of HSL on CMU-MOSI, CMU-MOSEI, and CH-SIMS under intra-modal missingness. $w/o$ (H,C) removes HSD and CM-CLSR. Results are averaged over $r \in \{0.0,0.1,\ldots,0.9\}$.}
\centering
\setlength{\tabcolsep}{3.3pt}
\renewcommand{\arraystretch}{1.0}
\begin{tabular}{l|cc|cc|cc}
\toprule
\multirow{2}{*}{\textbf{Model Config.}} & \multicolumn{2}{c|}{\textbf{CMU-MOSI}} & \multicolumn{2}{c|}{\textbf{CMU-MOSEI}} & \multicolumn{2}{c}{\textbf{CH-SIMS}} \\
\cline{2-7}
 & \textbf{F1(\%)}  & \textbf{MAE} $\downarrow$ & \textbf{F1(\%)}& \textbf{MAE} $\downarrow$ & \textbf{F1(\%)} & \textbf{MAE} $\downarrow$ \\
\midrule
\textbf{HSL (Ours)} & \textbf{\textcolor{red}{73.77}} & \textbf{0.965} & \textbf{81.45} & \textbf{0.635} & \textbf{\textcolor{red}{75.26}} & \textbf{0.476} \\
\midrule
\multicolumn{7}{c}{\textbf{(1) Importance of HSL Components}} \\
$w/o$ HSD & 71.24 & 1.012 & 78.56 & 0.672 & 72.39 & 0.510 \\
$w/o$ CM-CLSR & 68.95 & 1.082 & 76.82 & 0.715 & 70.06 & 0.552 \\
$w/o$ (H,C) & 66.54 & 1.134 & 74.36 & 0.735 & 68.01 & 0.586 \\
$w/o$ HCES & 72.58 & 0.995 & 79.45 & 0.658 & 73.11 & 0.495 \\
\midrule
\multicolumn{7}{c}{\textbf{(2) Effectiveness of Reconstruction Strategy}} \\
Bottom-Up & 71.55 & 1.020 & 78.65 & 0.675 & 72.51 & 0.515 \\
Single-Stage & 72.10 & 0.998 & 79.12 & 0.662 & 72.84& 0.505 \\
\midrule
\multicolumn{7}{c}{\textbf{(3) Different Fusion Mechanisms}} \\
Addition & 64.25 & 1.150 & 73.50 & 0.760 & 67.71 & 0.585 \\
Multiplication & 63.80 & 1.165 & 73.10 & 0.775 & 67.36 & 0.592 \\
TFN~\cite{TFN} & 70.85 & 1.035 & 77.95 & 0.685 & 71.16 & 0.535 \\
LMF~\cite{LMF} & 70.50 & 1.042 & 77.56 & 0.692 & 70.81 & 0.542 \\
CMAF~\cite{FDMER} & 73.15 & 0.985 & 80.25 & 0.648 & 74.02 & 0.496 \\
\midrule
\multicolumn{7}{c}{\textbf{(4) Evaluation of Regularization}} \\
$w/o$ $\mathcal{L}_{\text{rec}}$ & 69.10 & 1.075 & 77.05 & 0.710 & 70.36 & 0.548 \\
\bottomrule
\multicolumn{7}{l}{\footnotesize $\uparrow$ indicates higher is better and $\downarrow$ indicates lower is better.}
\end{tabular}
\label{tab:ablation_hsl_missingness}
\end{table}

\begin{figure}[t]
\centering
\includegraphics[width=0.95\linewidth]{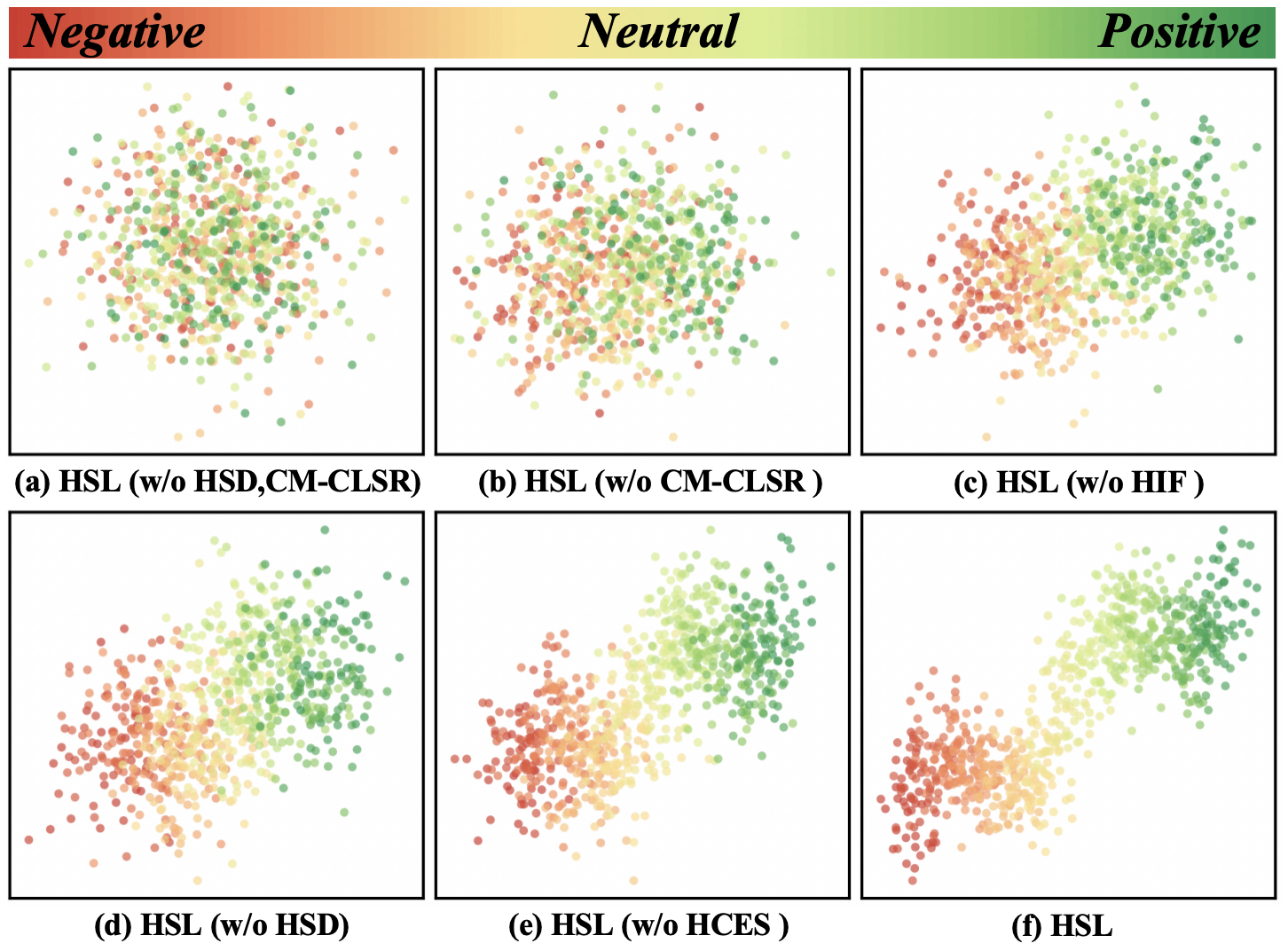}
\caption{Representation space visualization on CMU-MOSI. Compared with ablated variants, the full HSL model forms a more coherent feature manifold with a smoother sentiment gradient, consistent with improved semantic consistency under modality incompleteness.}
\label{fig:tsne}
\end{figure}

\subsubsection{Importance of HSL Components}
The first block of Table~\ref{tab:ablation_hsl_missingness} evaluates the contribution of each module.
Removing HSD degrades performance, which suggests that exposing hierarchical semantics helps reduce semantic averaging under missingness.
Removing CM-CLSR causes a larger drop, indicating that active reconstruction is critical when inputs contain erased tokens or frames.
Removing both HSD and CM-CLSR yields the lowest performance, which shows that hierarchical representation and hierarchical reconstruction are complementary.
Removing HCES also reduces performance, implying that prototype-based supervision helps prevent semantic collapse during the reconstruction cascade.

\subsubsection{Effectiveness of Reconstruction Strategy}
We compare the proposed top-down reconstruction with a bottom-up variant and a single-stage variant.
HSL outperforms the bottom-up variant, which supports the design choice that high-level semantics provide more modality-invariant anchors.
HSL also outperforms the single-stage variant, suggesting that progressive reconstruction simplifies the inverse problem by reducing ambiguity at each level.

\subsubsection{Different Fusion Mechanisms}
We compare HIF with alternative fusion strategies.
Simple additive and multiplicative fusion performs poorly since it cannot capture structured cross-modal and cross-level interactions.
Tensor-based fusion improves results but is limited in modeling hierarchical token sequences.
\textcolor{red}{The Cross-Modal Attention Fusion (CMAF) module of FDMER~\cite{FDMER} is competitive}, yet HIF yields stronger performance across datasets, which is consistent with its design that enables high-level semantics to route and refine lower-level cues through affinity-driven interactions.

\subsubsection{Evaluation of Regularization}
Removing $\mathcal{L}_{\text{rec}}$ leads to a clear degradation, which indicates that explicit reconstruction supervision is necessary for recovering missing distributions.
\textcolor{red}{Removing HCES ($w/o$ HCES)} also harms performance, showing that semantic-level constraints complement feature-level reconstruction by preventing reconstructed features from drifting away from correct semantic prototypes.

% \begin{figure}[ht]
%     \centering
%     \includegraphics[width=\linewidth]{weight-2.png}
%     \caption{Quantitative Analysis of Component Contribution on CMU-MOSI and CH-SIMS. (a) shows component weights measured by normalized gradient magnitudes during training. (b) shows component importance measured by performance drops of retrained ablation variants. The consistency between (a) and (b) indicates that training dynamics align with component level utility.}
%     \label{fig:weight}
% \end{figure}

\begin{figure}[ht]
    \centering
    \includegraphics[width=\linewidth]{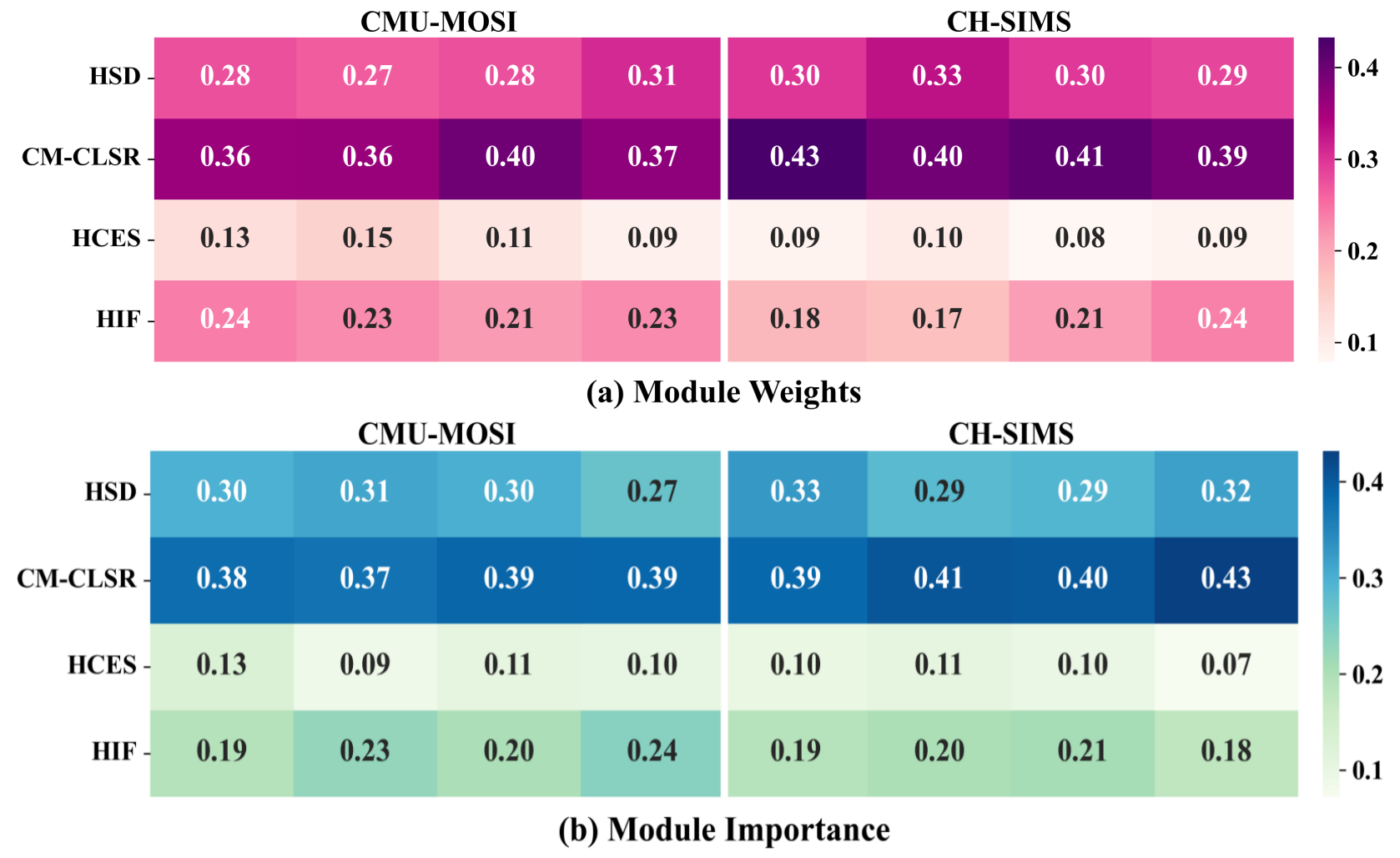}
    \caption{Quantitative analysis of component contribution on CMU-MOSI and CH-SIMS. (a) shows the component-wise normalized gradient contribution $w_u$ computed from parameter-scale-adjusted gradient intensities during training. (b) shows the normalized component importance $\bar{\Delta}_u$ computed from performance drops of retrained ablation variants. The consistency between (a) and (b) indicates that the training dynamics align with component-level utility.}
    \label{fig:weight}
\end{figure}

\subsection{Further Analysis}

% We provide further analyses to investigate representation geometry, component contributions, hyperparameter sensitivity, optimization stability, and the impact of hierarchical reconstruction.

We further analyze representation geometry, component contribution, hyperparameter sensitivity, optimization stability, and the effect of hierarchical reconstruction.

\subsubsection{Representation Space Visualization}
To qualitatively examine how HSL preserves representation geometry under modality incompleteness, we visualize the final fused features $h$ (defined in Section~\ref{sec:hif}) from the CMU-MOSI test set using t-SNE~\cite{t-sne}. Fig.~\ref{fig:tsne} compares the full HSL model against five ablation variants (four from Table~\ref{tab:ablation_hsl_missingness} plus $w/o$ HIF), with colors representing sentiment intensity. The variant lacking both HSD and CM-CLSR exhibits significant overlap between sentiment classes, indicating diminished separability. Removing individual components results in intermediate degradation, such as looser clusters or less coherent class boundaries. Conversely, the full HSL model produces a more compact distribution with a distinct monotonic transition across sentiment intensities. These visualizations corroborate our quantitative results, confirming that HSL’s hierarchical restoration maintains discriminative structural information despite missing inputs.

\subsubsection{Quantitative Analysis of Component Contribution}
\label{sec:interpretability}

We quantify component contributions using two complementary measures and examine their agreement on CMU-MOSI and CH-SIMS in Fig.~\ref{fig:weight}. The first measure is a gradient-based proxy that summarizes how strongly the training dynamics depend on each component. The second measure is an ablation-based importance score computed from retrained performance drops. 
Let $\mathcal{U}=\{\mathrm{HSD}, \mathrm{CM\text{-}CLSR}, \mathrm{HIF}, \mathrm{HCES}\}$ denote the set of analyzed components.
Let $\theta_u$ denote the parameters of component $u \in \mathcal{U}$ and let $\mathcal{L}_{\mathrm{total}}$ be the training objective in Section~\ref{sec:objective}. \textcolor{red}{Within each dataset, the four columns correspond to four sample batches. Each column reports the normalized contributions of the four components within the corresponding batch.} We first compute a parameter-scale-adjusted gradient intensity $g_u$ to reduce the influence of different component sizes, where $|\theta_u|$ is the number of parameters in component $u$ and the expectation is taken over training mini-batches across the training procedure.
We then normalize across components to obtain $w_u$, which provides a comparable proxy of optimization reliance among components.
\begin{equation}
g_u = \mathbb{E}\left[\frac{\|\nabla_{\theta_u}\mathcal{L}_{\mathrm{total}}\|_2}{\sqrt{|\theta_u|}}\right],
w_u = \frac{g_u}{\sum_{u' \in \mathcal{U}} g_{u'}},
\end{equation}
Therefore, Fig.~\ref{fig:weight}(a) visualizes the component-wise normalized gradient contribution $w_u$, rather than the raw gradient magnitude or the unnormalized intensity $g_u$.
This gradient-based measure is used as a relative diagnostic of optimization reliance and is interpreted together with the retrained ablation drops below.
To estimate component importance, for each $u \in \mathcal{U}$ we retrain an ablated variant $\mathrm{HSL}^{(-u)}$ that removes $u$ while keeping all other settings identical, and we evaluate performance under the same protocol as the full model. Using a higher-is-better primary metric, we define the retrained performance drop as $\Delta_u = \mathrm{Perf}(\mathrm{HSL}) - \mathrm{Perf}(\mathrm{HSL}^{(-u)})$. For visualization, we normalize drops by their sum $\bar{\Delta}_u = \frac{\Delta_u}{\sum_{u' \in \mathcal{U}} \Delta_{u'}}$. 
Thus, Fig.~\ref{fig:weight}(b) reports the normalized ablation-based importance $\bar{\Delta}_u$.
For both CMU-MOSI and CH-SIMS, we use the F1 as $\mathrm{Perf}(\cdot)$, following the same definition reported in the main robustness tables. Across both datasets, CM\text{-}CLSR exhibits the largest $w_u$ and the largest $\bar{\Delta}_u$, which is consistent with Table~\ref{tab:ablation_hsl_missingness} and supports that cross-level reconstruction is central to robustness under missing modalities. HSD is the second most influential component, reflecting the benefit of exposing a semantic hierarchy before reconstruction and fusion. HIF contributes stably as the fusion backbone. HCES shows a smaller but consistent contribution, consistent with its role as an auxiliary semantic regularizer during training. Overall, the agreement between $w_u$ and $\bar{\Delta}_u$ suggests that the optimization emphasis is aligned with the measured utility of each component.

\begin{figure}[t]
    \centering
    \includegraphics[width=\linewidth]{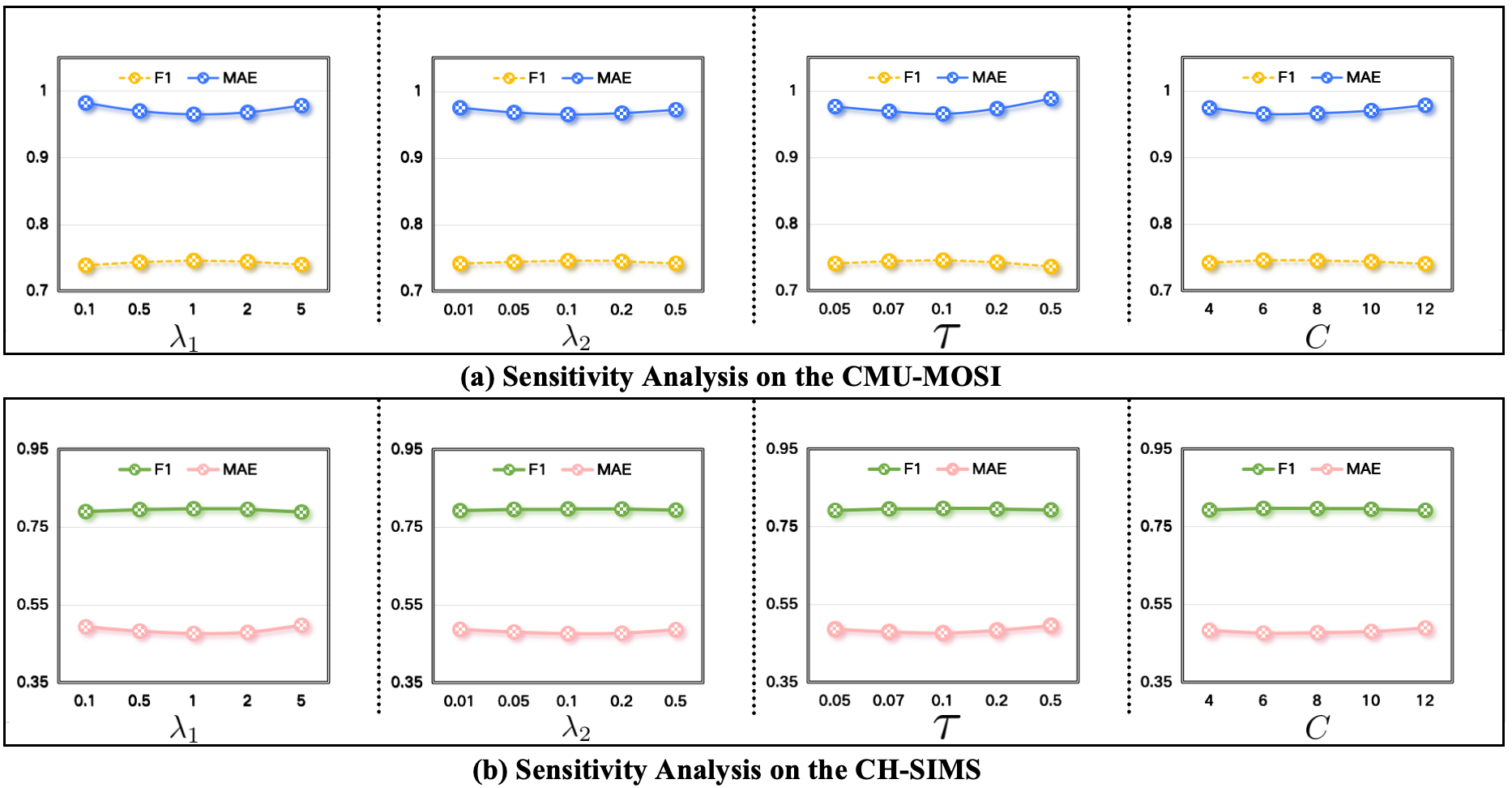}
    \caption{Sensitivity Analysis on benchmarks under intra-modal missingness.}
    \label{fig:Sensitivity}
\end{figure}

\subsubsection{Sensitivity Analysis}
We evaluate the sensitivity of HSL to four key hyperparameters: the reconstruction loss weight $\lambda_1$, the HCES regularization weight $\lambda_2$, the temperature $\tau$, and the prototype count $C$. Fig.~\ref{fig:Sensitivity} reports one-at-a-time performance variations by changing a single hyperparameter while fixing the others to the default configuration. Overall, HSL maintains stable MAE and F1 over a wide range around the chosen configuration, indicating that the proposed objectives and constraints do not rely on fragile hyperparameter tuning. In particular, performance remains competitive across moderate ranges of $\lambda_1$ and $\lambda_2$, supporting that reconstruction and semantic supervision complement the task objective without destabilizing optimization. For prototype learning, an intermediate $\tau$ yields the best trade-off between discriminability and optimization stability, while $C$ shows limited sensitivity within a reasonable range, suggesting that the hierarchical design captures semantic structure beyond a particular discretization granularity.

\begin{figure}[t]
    \centering
    \includegraphics[width=\linewidth]{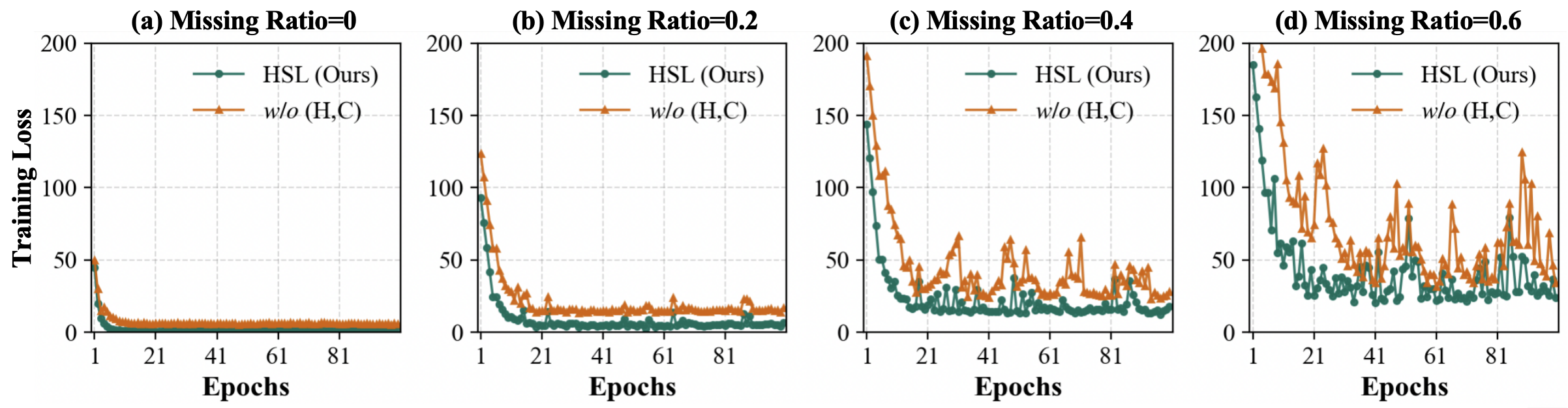}
    \caption{Optimization stability under different missing rates $r$. HSL shows smoother training dynamics than the variant without HSD and CM-CLSR.}
    \label{fig:adv}
\end{figure}

\subsubsection{Advantages of the Hierarchical Semantic Learning Design}
We analyze optimization stability by comparing training loss trajectories of HSL with the degenerated variant that removes both HSD and CM-CLSR. Experiments are conducted under multiple missing rates $r \in \{0.0, 0.2, 0.4, 0.6\}$. As shown in Fig.~\ref{fig:adv}, the degenerated variant exhibits larger oscillations and slower convergence as missingness increases, which is consistent with the difficulty of learning from flat representations with zero padding. In contrast, HSL presents smoother loss descent across missing rates, suggesting that hierarchical exposure and top-down reconstruction help reduce distribution shift induced by missing data and yield more stable learning dynamics.

\begin{figure}[t]
    \centering
    \includegraphics[width=\linewidth]{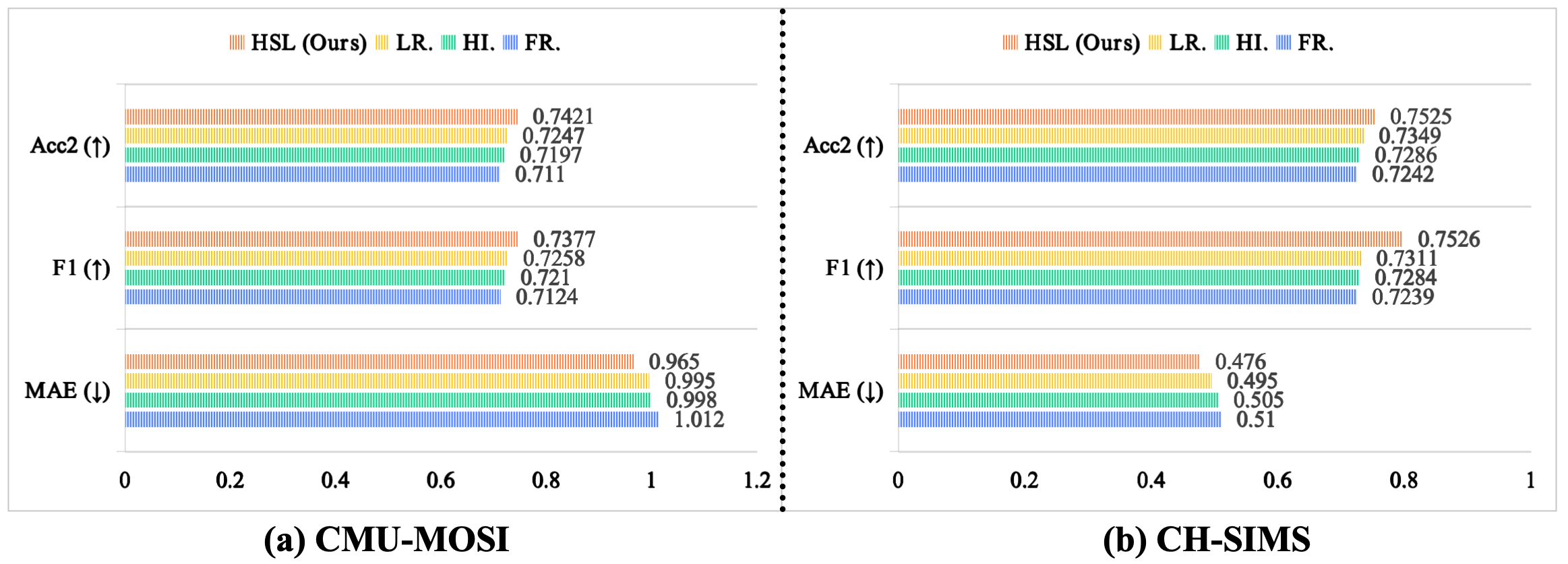}
    \caption{\textcolor{red}{Necessity analysis of hierarchical reconstruction paradigms on CMU-MOSI and CH-SIMS under intra-modal missingness, averaged over $r \in \{0.0,0.1,\ldots,0.9\}$. FR, HI, and LR correspond to the w/o HSD, Single-Stage, and \textit{w/o} HCES variants in Table~\ref{tab:ablation_hsl_missingness}.}}
    \label{fig:NHR}
\end{figure}

\subsubsection{Necessity Analysis of Hierarchical Reconstruction}
We examine the hypothesis that reconstructing only a single flattened representation is insufficient for robust learning under missing modalities. This setting is challenging because low-level cues are largely modality-specific and cross-modal correspondences at this scale are weak, which makes the completion problem underdetermined. Under the intra-modal missingness protocol of Table~\ref{tab:ablation_hsl_missingness}, we compare four reconstruction paradigms. Flattened Reconstruction (FR) reconstructs only the final-stage representation and performs fusion at that stage (the $w/o$ HSD variant in Table~\ref{tab:ablation_hsl_missingness}). Hierarchical Independent Reconstruction (HI) exposes three semantic levels and reconstructs each level independently without conditioning on reconstructed higher-level semantics (the Single-Stage variant). Ladder Reconstruction (LR) performs progressive top-down reconstruction across levels but removes HCES and therefore lacks level-wise semantic supervision (the $w/o$ HCES variant). Full HSL combines progressive reconstruction with HCES. As shown in Fig.~\ref{fig:NHR}, exposing a hierarchy improves over flattened reconstruction, indicating that separating semantic levels helps constrain reconstruction. Ladder reconstruction yields further gains, indicating that high-level semantics provide effective priors for mid- and low-level recovery. Full HSL performs best, which is consistent with the role of level-wise supervision in reducing error accumulation during progressive refinement.

\begin{figure}[ht]
    \centering
    \includegraphics[width=\linewidth]{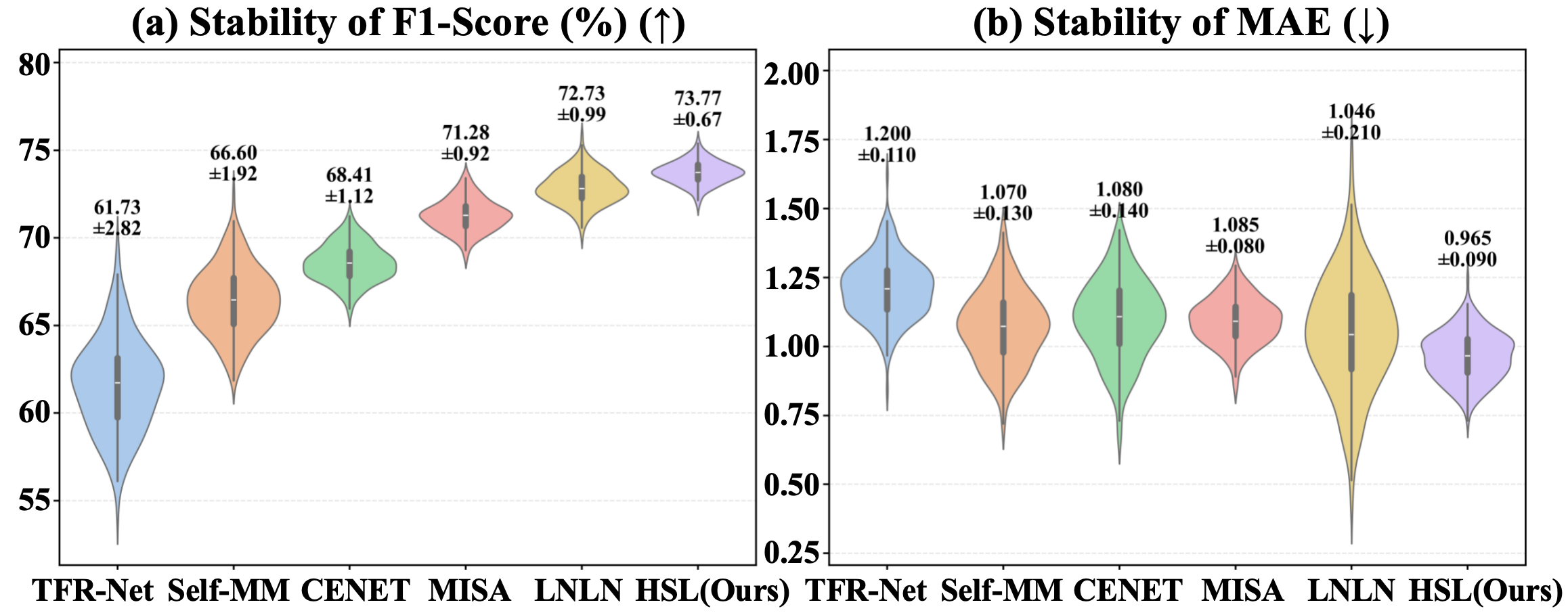}
    \caption{\textcolor{red}{Performance variation across missing rates and seed stability on CMU-MOSI. Each violin shows ten per-rate means (\(r\in\{0.0,0.1,\ldots,0.9\}\)). HSL statistics are computed over five random seeds, whereas baseline statistics are taken from LNLN~\cite{LNLN} and follow its reported three-seed evaluation. The annotated mean is averaged across missing rates, while ± denotes the average within-rate standard deviation across seeds. Violin spread reflects variation across missing rates.}}
    \label{fig:Stability}
\end{figure}

\subsubsection{Statistical Stability Analysis}
\label{sec:Statistical Stability Analysis}
\textcolor{red}{To assess performance variation across missing rates and stability across random initializations, we compare HSL with representative baselines on CMU-MOSI in Fig.~\ref{fig:Stability}. Each violin contains ten per-rate mean scores, with each score averaged over random seeds. The annotated standard deviation is the average of the within-rate standard deviations across seeds, rather than the standard deviation of the violin distribution. Thus, the violin captures variation across missing rates, whereas the reported standard deviation reflects seed-level variability. HSL statistics are computed over five seeds, while baseline statistics are taken from LNLN~\cite{LNLN}. HSL achieves the highest average Non0 F1 and the lowest average MAE among the compared methods, with seed-level variability comparable to or smaller than that of the baselines.}

Moreover, under the same experimental protocol, HSL has the shortest training time among the methods compared in Appendix E (17.76 s per epoch, vs. 18.67 s for P-RMF, and 24.66 s for LNLN) with a comparable number of parameters. Details of the efficiency analysis are provided in Appendix E.

\begin{figure}[ht]
    \centering
    \includegraphics[width=\linewidth]{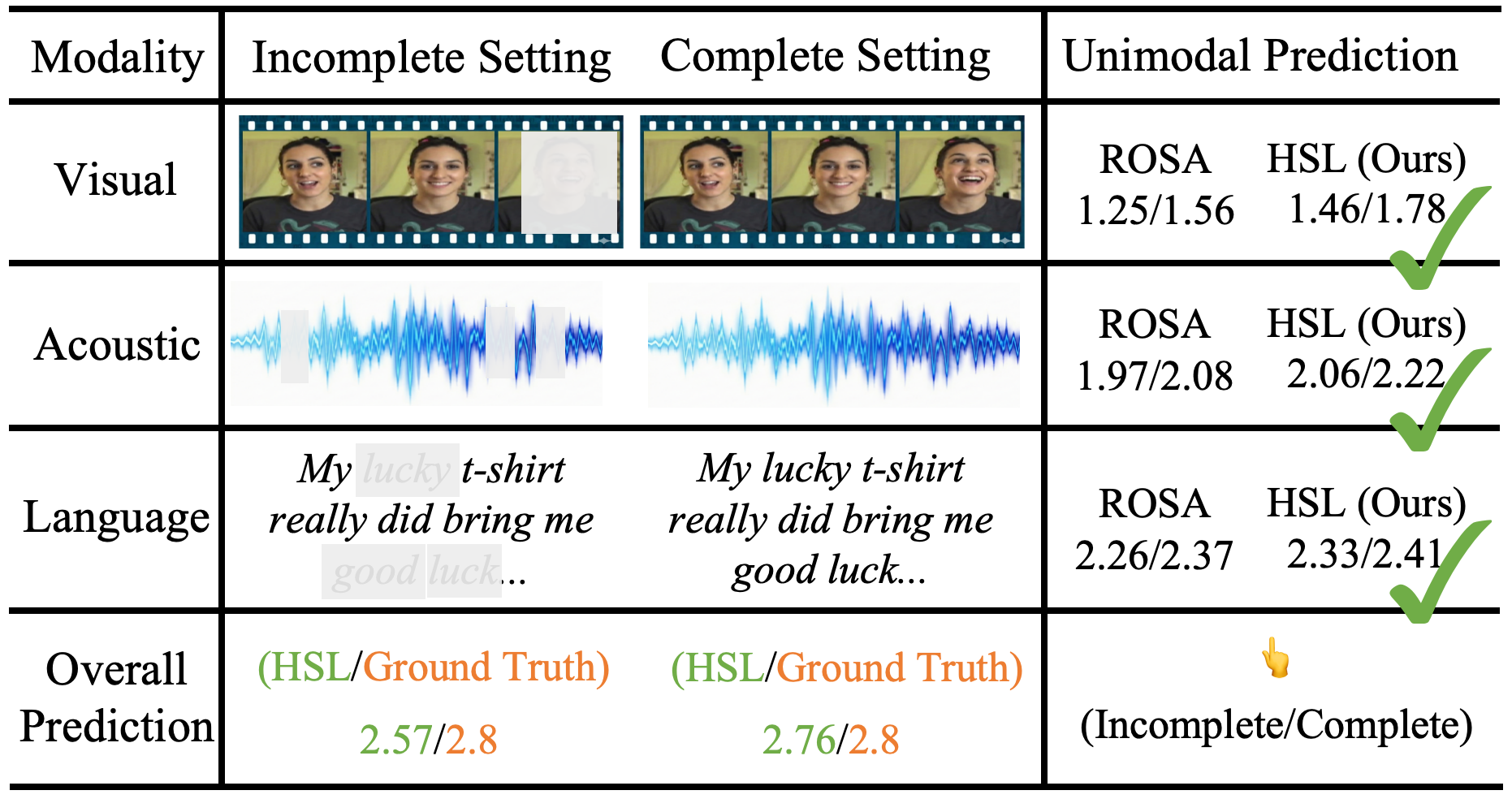}
    \caption{Case study under intra-modal missingness with missing rate $0.3$. Unimodal predictions are reported as (Incomplete/Complete), and the Overall Prediction row shows HSL's fused prediction versus the ground truth.}
    \label{fig:Case_Study}
\end{figure}

\subsection{Case Study}

Fig.~\ref{fig:Case_Study} illustrates an example under intra-modal missingness with missing rate $0.3$.
In the incomplete setting, 30\% of tokens or frames are randomly masked within each modality stream, while the complete setting uses the original inputs.
For each modality, we show unimodal sentiment predictions of ROSA and HSL, reported as (Incomplete/Complete). We also report the fused prediction of HSL and the ground-truth sentiment score. Under incomplete inputs, HSL produces coherent unimodal predictions across visual, acoustic, and language streams, and its fused prediction remains close to the ground truth. In this example, HSL outputs 2.57 in the incomplete setting and 2.76 in the complete setting, while the ground-truth score is 2.8. This case supports that reconstructing semantics in a level-wise manner can compensate for missing local evidence and stabilizes the final fusion outcome under intra-modal corruption.

% \section{Discussion}

% Although our model has advantages in the MSA task, there are still some limitations. Firstly, due to the introduction of a large language model and Whisper model, HSL requires greater
% computational resources compared to other “small” models, but this is a necessary cost for leveraging the capabilities of the large model. Secondly, the method of uniformly sampling 8 frames
% is suitable for short videos. For long videos, more effective methods of frame extraction and information compression are necessary. Finally, our model takes into account the three most commonly used modalities, but needs to adjust the model architecture when other modalities such as brain waves are added.

\section{Limitations and Future Work} 

Although HSL improves the learning of multimodal representations under missing modalities, several limitations remain. In this study, we instantiate and evaluate HSL in the commonly used trimodal setting (language, acoustic, and visual). In real-world deployments, however, systems may involve richer and more heterogeneous modality sets (e.g., infrared, LiDAR) with diverse temporal/spatial structures and sampling rates. An important direction for future work is to extend HSL to such settings by designing modality-aware interface adaptations while preserving its core hierarchical reconstruction principle. 

\textbf{Generalization to heterogeneous sensor modalities.}
The current HSL uses a three-level hierarchy for trimodal MSA with language, acoustic, and visual modalities. This should be viewed as a practical instantiation for deep modality-specific encoders, rather than a universal assumption that all sensor modalities require exactly three stages. For wild heterogeneous sensors, such as PPG or HRV, the proper hierarchy depth may vary with signal type, encoder depth, sampling frequency, temporal granularity, and task abstraction. HRV may only need a coarse local-to-global physiological hierarchy, whereas high-frequency PPG may require finer levels covering waveform morphology, beat-level patterns, segment-level dynamics, and task-level physiological semantics.

HSL can therefore be generalized from $\mathcal{M}=\{l,a,v\}$ to $\mathcal{M}=\{m_1,\ldots,m_N\}$, and from a fixed three-level hierarchy to a shared $K$-level hierarchy or modality-specific hierarchies $\{K_m\}_{m\in\mathcal{M}}$. Its essential principle is to preserve internal semantic hierarchy and use higher-level semantics to constrain lower-level reconstruction, rather than to fix the number of levels. When modalities adopt different hierarchy depths, level projection or semantic-scale matching may be required before cross-level reconstruction and hierarchical fusion. We leave automatic hierarchy-depth selection and validation on wild sensor data as future work.

\section{Conclusion}
In this study, we propose Hierarchical Semantic Learning (HSL), a hierarchical reconstruction method for MSA with missing modalities. Rather than relying solely on final-layer representations, HSL exposes a three-level semantic hierarchy for each modality and conducts top-down, anchor-guided reconstruction from high-level semantics to mid- and low-level details. Specifically, HSD makes the depth-wise semantic progression explicit, CM-CLSR reconstructs missing modality features via a high-level anchor and semantic injection, and HCES applies level-wise prototype-based supervision to regularize reconstructed features and mitigate error accumulation along the reconstruction cascade. Finally, HIF models cross-modal and cross-level dependencies to produce a task-ready fused representation. Experiments on four benchmarks under intra-modal and fixed-modality missingness show consistent improvements over representative baselines \textcolor{red}{on average and in most settings}, and ablations and analyses further corroborate the contribution of each component.

\bibliographystyle{IEEEtran}
\bibliography{TMM-References}

@inproceedings{misa,
  author    = {Hazarika, Devamanyu and Zimmermann, Roger and Poria, Soujanya},
  title     = {{MISA: Modality-Invariant and -Specific Representations for Multimodal Sentiment Analysis}},
  booktitle = {Proceedings of the 28th ACM International Conference on Multimedia},
  year      = {2020},
  pages     = {1122--1131},
  doi       = {10.1145/3394171.3413678}
}

@inproceedings{FDMER,
  author    = {Yang, Dingkang and Huang, Shuai and Kuang, Haopeng and Du, Yangtao and Zhang, Lihua},
  title     = {{Disentangled Representation Learning for Multimodal Emotion Recognition}},
  booktitle = {Proceedings of the 30th ACM International Conference on Multimedia},
  year      = {2022},
  pages     = {1642--1651},
  doi       = {10.1145/3503161.3547754}
}

@article{ROSA,
  author  = {Li, Ziming and Liu, Yaxin and Yang, Chuanpeng and Zhou, Yan and Hu, Songlin},
  title   = {{ROSA: A Robust Self-Adaptive Model for Multimodal Emotion Recognition with Uncertain Missing Modalities}},
  journal = {IEEE Transactions on Multimedia},
  year    = {2025},
  volume  = {27},
  pages   = {6766--6779},
  doi     = {10.1109/TMM.2025.3590929}
}

@article{MLCL,
  author  = {Zhuang, Yan and Bai, Wei and Zhang, Yanru and Deng, Jiawen and Hu, Zheng and Zhang, Xiaoyue and Ren, Fuji},
  title   = {{Multi-Level Contrastive Learning for Multimodal Sentiment Analysis}},
  journal = {IEEE Transactions on Multimedia},
  year    = {2025},
  volume  = {27},
  pages   = {9044--9058},
  doi     = {10.1109/TMM.2025.3613116}
}

@inproceedings{DAR,
  author    = {Yang, Mingzheng and Zhang, Kai and Ye, Yuyang and Zhang, Yanghai and Yu, Runlong and Hou, Min},
  title     = {{Decoupling and Reconstructing: A Multimodal Sentiment Analysis Framework Towards Robustness}},
  booktitle = {Proceedings of the Thirty-Fourth International Joint Conference on Artificial Intelligence},
  year      = {2025},
  pages     = {6803--6811},
  doi       = {10.24963/ijcai.2025/757}
}

@inproceedings{LNLN,
  author    = {Zhang, Haoyu and Wang, Wenbin and Yu, Tianshu},
  title     = {{Towards Robust Multimodal Sentiment Analysis with Incomplete Data}},
  booktitle = {Advances in Neural Information Processing Systems},
  volume    = {37},
  year      = {2024},
  pages     = {55943--55974},
  doi       = {10.52202/079017-1779}
}

@inproceedings{P-RMF,
  author    = {Zhu, Aoqiang and Hu, Min and Wang, Xiaohua and Yang, Jiaoyun and Tang, Yiming and An, Ning},
  title     = {{Proxy-Driven Robust Multimodal Sentiment Analysis with Incomplete Data}},
  booktitle = {Proceedings of the 63rd Annual Meeting of the Association for Computational Linguistics (Volume 1: Long Papers)},
  year      = {2025},
  pages     = {22123--22138},
  doi       = {10.18653/v1/2025.acl-long.1075},
  url       = {https://aclanthology.org/2025.acl-long.1075/}
}

@article{HKD-MER,
  author  = {Sun, Teng and Wei, Yinwei and Ni, Juntong and Liu, Zixin and Song, Xuemeng and Wang, Yaowei and Nie, Liqiang},
  title   = {{Muti-modal Emotion Recognition via Hierarchical Knowledge Distillation}},
  journal = {IEEE Transactions on Multimedia},
  year    = {2024},
  volume  = {26},
  pages   = {9036--9046},
  doi     = {10.1109/TMM.2024.3385180}
}

@inproceedings{IMDer,
  author    = {Wang, Yuanzhi and Li, Yong and Cui, Zhen},
  title     = {{Incomplete Multimodality-Diffused Emotion Recognition}},
  booktitle = {Advances in Neural Information Processing Systems},
  volume    = {36},
  year      = {2023},
  pages     = {17117--17128},
  doi       = {10.52202/075280-0748}
}

@article{GCNet,
  author  = {Lian, Zheng and Chen, Lan and Sun, Licai and Liu, Bin and Tao, Jianhua},
  title   = {{GCNet: Graph Completion Network for Incomplete Multimodal Learning in Conversation}},
  journal = {IEEE Transactions on Pattern Analysis and Machine Intelligence},
  year    = {2023},
  volume  = {45},
  number  = {7},
  pages   = {8419--8432},
  doi     = {10.1109/TPAMI.2023.3234553}
}

@article{EMT,
  author  = {Sun, Licai and Lian, Zheng and Liu, Bin and Tao, Jianhua},
  title   = {{Efficient Multimodal Transformer with Dual-Level Feature Restoration for Robust Multimodal Sentiment Analysis}},
  journal = {IEEE Transactions on Affective Computing},
  year    = {2024},
  volume  = {15},
  number  = {1},
  pages   = {309--325},
  doi     = {10.1109/TAFFC.2023.3274829}
}

@article{MIG-HCL,
  author  = {Gong, Peizhu and Liu, Jin and Zhang, Xiliang and Li, Xingye and Wei, Lai and He, Huihua},
  title   = {{Towards Robust Sentiment Analysis with Multimodal Interaction Graph and Hybrid Contrastive Learning}},
  journal = {Pattern Recognition},
  year    = {2026},
  volume  = {169},
  pages   = {111870},
  doi     = {10.1016/j.patcog.2025.111870},
  note    = {Article number 111870}
}

@inproceedings{ALMT,
  author    = {Zhang, Haoyu and Wang, Yu and Yin, Guanghao and Liu, Kejun and Liu, Yuanyuan and Yu, Tianshu},
  title     = {{Learning Language-Guided Adaptive Hyper-Modality Representation for Multimodal Sentiment Analysis}},
  booktitle = {Proceedings of the 2023 Conference on Empirical Methods in Natural Language Processing},
  year      = {2023},
  pages     = {756--767},
  doi       = {10.18653/v1/2023.emnlp-main.49},
  url       = {https://aclanthology.org/2023.emnlp-main.49/}
}

@article{NIAT,
  author  = {Yuan, Ziqi and Liu, Yihe and Xu, Hua and Gao, Kai},
  title   = {{Noise Imitation Based Adversarial Training for Robust Multimodal Sentiment Analysis}},
  journal = {IEEE Transactions on Multimedia},
  year    = {2024},
  volume  = {26},
  pages   = {529--539},
  doi     = {10.1109/TMM.2023.3267882}
}

@inproceedings{TSDA,
  author    = {Meng, Chunlei and Zhou, Ziyang and He, Lucas and Du, Xiaojing and Ouyang, Chun and Gan, Zhongxue},
  title     = {{Temporal-Spatial Decouple Before Act: Disentangled Representation Learning for Multimodal Sentiment Analysis}},
  booktitle = {2026 IEEE International Conference on Acoustics, Speech and Signal Processing (ICASSP)},
  year      = {2026},
  pages     = {11917--11921},
  doi       = {10.1109/ICASSP55912.2026.11463049}
}

@inproceedings{EAU,
  author    = {Gao, Zixian and Jiang, Xun and Xu, Xing and Shen, Fumin and Li, Yujie and Shen, Heng Tao},
  title     = {{Embracing Unimodal Aleatoric Uncertainty for Robust Multimodal Fusion}},
  booktitle = {Proceedings of the IEEE/CVF Conference on Computer Vision and Pattern Recognition},
  year      = {2024},
  pages     = {26876--26885},
  doi       = {10.1109/CVPR52733.2024.02538}
}

@inproceedings{dmd,
  author    = {Li, Yong and Wang, Yuanzhi and Cui, Zhen},
  title     = {{Decoupled Multimodal Distilling for Emotion Recognition}},
  booktitle = {Proceedings of the IEEE/CVF Conference on Computer Vision and Pattern Recognition},
  year      = {2023},
  pages     = {6631--6640}
}

@inproceedings{2019BERT,
  author    = {Jawahar, Ganesh and Sagot, Beno{\^i}t and Seddah, Djam{\'e}},
  title     = {What Does {BERT} Learn about the Structure of Language?},
  booktitle = {Proceedings of the 57th Annual Meeting of the Association for Computational Linguistics},
  year      = {2019},
  pages     = {3651--3657},
  doi       = {10.18653/v1/P19-1356},
  url       = {https://aclanthology.org/P19-1356/}
}

@inproceedings{pix2pix,
  author    = {Isola, Phillip and Zhu, Jun-Yan and Zhou, Tinghui and Efros, Alexei A.},
  title     = {{Image-to-Image Translation with Conditional Adversarial Networks}},
  booktitle = {2017 IEEE Conference on Computer Vision and Pattern Recognition (CVPR)},
  year      = {2017},
  pages     = {5967--5976},
  doi       = {10.1109/CVPR.2017.632}
}

@inproceedings{SRGAN,
  author    = {Ledig, Christian and Theis, Lucas and Husz{\'a}r, Ferenc and Caballero, Jose and Cunningham, Andrew and Acosta, Alejandro and Aitken, Andrew and Tejani, Alykhan and Totz, Johannes and Wang, Zehan and Shi, Wenzhe},
  title     = {{Photo-Realistic Single Image Super-Resolution Using a Generative Adversarial Network}},
  booktitle = {Proceedings of the IEEE Conference on Computer Vision and Pattern Recognition},
  year      = {2017},
  pages     = {4681--4690},
  doi       = {10.1109/CVPR.2017.19}
}

@article{UniMF,
  author  = {Huan, Ruohong and Zhong, Guowei and Chen, Peng and Liang, Ronghua},
  title   = {{UniMF: A Unified Multimodal Framework for Multimodal Sentiment Analysis in Missing Modalities and Unaligned Multimodal Sequences}},
  journal = {IEEE Transactions on Multimedia},
  year    = {2024},
  volume  = {26},
  pages   = {5753--5768},
  doi     = {10.1109/TMM.2023.3338769}
}

@article{TagEnc,
  author  = {Zeng, Jiandian and Zhou, Jiantao and Liu, Tianyi},
  title   = {{Robust Multimodal Sentiment Analysis via Tag Encoding of Uncertain Missing Modalities}},
  journal = {IEEE Transactions on Multimedia},
  year    = {2023},
  volume  = {25},
  pages   = {6301--6314},
  doi     = {10.1109/TMM.2022.3207572}
}

@article{MissModal,
  author  = {Lin, Ronghao and Hu, Haifeng},
  title   = {{MissModal: Increasing Robustness to Missing Modality in Multimodal Sentiment Analysis}},
  journal = {Transactions of the Association for Computational Linguistics},
  year    = {2023},
  volume  = {11},
  pages   = {1686--1702},
  doi     = {10.1162/tacl_a_00628},
  url     = {https://aclanthology.org/2023.tacl-1.94/}
}

@inproceedings{DCCA,
  author    = {Andrew, Galen and Arora, Raman and Bilmes, Jeff and Livescu, Karen},
  title     = {{Deep Canonical Correlation Analysis}},
  booktitle = {Proceedings of the 30th International Conference on Machine Learning},
  series    = {Proceedings of Machine Learning Research},
  volume    = {28},
  number    = {3},
  year      = {2013},
  pages     = {1247--1255},
  publisher = {PMLR},
  url       = {https://proceedings.mlr.press/v28/andrew13.html}
}

@inproceedings{DCCAE,
  author    = {Wang, Weiran and Arora, Raman and Livescu, Karen and Bilmes, Jeff},
  title     = {{On Deep Multi-View Representation Learning}},
  booktitle = {Proceedings of the 32nd International Conference on Machine Learning},
  series    = {Proceedings of Machine Learning Research},
  volume    = {37},
  year      = {2015},
  pages     = {1083--1092},
  publisher = {PMLR},
  url       = {https://proceedings.mlr.press/v37/wangb15.html}
}

@InProceedings{TSD,
    author    = {Meng, Chunlei and Luo, Jiabin and Yan, Zhenglin and Yu, Zhenyu and Fu, Rong and Gan, Zhongxue and Ouyang, Chun},
    title     = {Tri-Subspaces Disentanglement for Multimodal Sentiment Analysis},
    booktitle = {Proceedings of the IEEE/CVF Conference on Computer Vision and Pattern Recognition (CVPR)},
    year      = {2026},
    pages     = {8791-8800},
    publisher = {IEEE/CVF}
}

@InProceedings{GCL,
  title =   {Group Cognition Learning: Making Everything Better Through Governed Two-Stage Agents Collaboration},
  author =       {Meng, Chunlei and Feng, Pengbin and Fu, Rong and Lee, Hoi Leong and Du, Xiaojing and Kang, Zhaolu and Zhang, Zeyu and Zhou, Weilin and Ouyang, Chun and Gan, Zhongxue},
  booktitle =   {Proceedings of the 43rd International Conference on Machine Learning},
  pages =   {87822--87834},
  year =   {2026},
  volume =   {306},
  series =   {Proceedings of Machine Learning Research},
  publisher =    {PMLR},
  url =   {https://proceedings.mlr.press/v306/meng26a.html}
}

@inproceedings{DBR,
  author = {Meng, Chunlei and Luo, Jiabin and Feng, Pengbin and Yan, Zhenglin and Hu, Chengyin and Gan, Zhongxue and Ouyang, Chun},
  title = {Mitigating Shared-Private Branch Imbalance via Dual-Branch Rebalancing for Multimodal Sentiment Analysis},
  isbn = {9798400722134},
  publisher = {Association for Computing Machinery},
  booktitle = {Proceedings of the 34th ACM International Conference on Multimedia},
  pages = {7690–7698},
  numpages = {9},
  year = {2026}
}

@InProceedings{CLCR,
    author    = {Meng, Chunlei and Huang, Guanhong and Fu, Rong and Jian, Runmin and Gan, Zhongxue and Ouyang, Chun},
    title     = {CLCR: Cross-Level Semantic Collaborative Representation for Multimodal Learning},
    booktitle = {Proceedings of the IEEE/CVF Conference on Computer Vision and Pattern Recognition (CVPR)},
    year      = {2026},
    pages     = {1606-1615},
    publisher = {IEEE/CVF}
}

@inproceedings{AHRA,
    title = {Adaptive Hierarchical Representation Alliance for Multimodal Learning},
    author = {Meng, Chunlei and Feng, Pengbin and Pang, Jacqueline J.  and Liao, Chih-Ting and Fu, Rong  and Kang, Zhaolu  and Gan, Zhongxue  and Ouyang, Chun},
    booktitle = {Findings of the {A}ssociation for {C}omputational {L}inguistics: {EMNLP} 2026},
    year = {2026},
    publisher = "Association for Computational Linguistics",
    url = "https://aclanthology.org/2026.findings-emnlp.1014/",
    pages = {21423--21439},
}

@misc{MRCF,
      title={Rethinking Modality Reliability in Multimodal Sentiment Analysis with Incomplete Observations},
      author={Chunlei Meng and Jacqueline J. Pang and Pengbin Feng and Zhenyu Yu and Chun Ouyang and Zhongxue Gan},
      year={2026},
      eprint={2608.03611},
      archivePrefix={arXiv},
      primaryClass={cs.AI},
      url={https://arxiv.org/abs/2608.03611}, 
}

@misc{PSM-WSM,
      title={PSM-WSM: Predictively Sufficient Multimodal World Social Modeling}, 
      author={Chunlei Meng and Zhenyu Yu and Pengbin Feng and Jacqueline J. Pang and Jiabin Luo and Chih-Ting Liao and Zhaolu Kang and Chun Ouyang and Zhongxue Gan},
      year={2026},
      eprint={2611.25835},
      archivePrefix={arXiv},
      primaryClass={cs.MM},
      url={https://arxiv.org/abs/2611.25835}, 
}

@inproceedings{MMIN,
  author    = {Zhao, Jinming and Li, Ruichen and Jin, Qin},
  title     = {{Missing Modality Imagination Network for Emotion Recognition with Uncertain Missing Modalities}},
  booktitle = {Proceedings of the 59th Annual Meeting of the Association for Computational Linguistics and the 11th International Joint Conference on Natural Language Processing (Volume 1: Long Papers)},
  year      = {2021},
  pages     = {2608--2618},
  doi       = {10.18653/v1/2021.acl-long.203},
  url       = {https://aclanthology.org/2021.acl-long.203/}
}

@inproceedings{GGMD,
  author    = {Wang, Hao and Luo, Shengda and Hu, Guosheng and Zhang, Jianguo},
  title     = {{Gradient-Guided Modality Decoupling for Missing-Modality Robustness}},
  booktitle = {Proceedings of the AAAI Conference on Artificial Intelligence},
  volume    = {38},
  number    = {14},
  year      = {2024},
  pages     = {15483--15491},
  doi       = {10.1609/AAAI.V38I14.29474}
}

@inproceedings{CorrKD,
  author    = {Li, Mingcheng and Yang, Dingkang and Zhao, Xiao and Wang, Shuaibing and Wang, Yan and Yang, Kun and Sun, Mingyang and Kou, Dongliang and Qian, Ziyun and Zhang, Lihua},
  title     = {{Correlation-Decoupled Knowledge Distillation for Multimodal Sentiment Analysis with Incomplete Modalities}},
  booktitle = {Proceedings of the IEEE/CVF Conference on Computer Vision and Pattern Recognition},
  year      = {2024},
  pages     = {12458--12468},
  doi       = {10.1109/CVPR52733.2024.01184}
}

@inproceedings{MMIM,
  author    = {Han, Wei and Chen, Hui and Poria, Soujanya},
  title     = {{Improving Multimodal Fusion with Hierarchical Mutual Information Maximization for Multimodal Sentiment Analysis}},
  booktitle = {Proceedings of the 2021 Conference on Empirical Methods in Natural Language Processing},
  year      = {2021},
  pages     = {9180--9192},
  doi       = {10.18653/v1/2021.emnlp-main.723},
  url       = {https://aclanthology.org/2021.emnlp-main.723/}
}

@article{CENET,
  author  = {Wang, Di and Liu, Shuai and Wang, Quan and Tian, Yumin and He, Lihuo and Gao, Xinbo},
  title   = {{Cross-Modal Enhancement Network for Multimodal Sentiment Analysis}},
  journal = {IEEE Transactions on Multimedia},
  year    = {2023},
  volume  = {25},
  pages   = {4909--4921},
  doi     = {10.1109/TMM.2022.3183830}
}

@article{CIF-MMIN,
  author  = {Liu, Rui and Zuo, Haolin and Lian, Zheng and Schuller, Bj{\"o}rn W. and Li, Haizhou},
  title   = {{Contrastive Learning Based Modality-Invariant Feature Acquisition for Robust Multimodal Emotion Recognition with Missing Modalities}},
  journal = {IEEE Transactions on Affective Computing},
  year    = {2024},
  volume  = {15},
  number  = {4},
  pages   = {1856--1873},
  doi     = {10.1109/TAFFC.2024.3378570}
}

@inproceedings{TFR-Net,
  author    = {Yuan, Ziqi and Li, Wei and Xu, Hua and Yu, Wenmeng},
  title     = {{Transformer-Based Feature Reconstruction Network for Robust Multimodal Sentiment Analysis}},
  booktitle = {Proceedings of the 29th ACM International Conference on Multimedia},
  year      = {2021},
  pages     = {4400--4407},
  doi       = {10.1145/3474085.3475585}
}

@inproceedings{MulT,
  author    = {Tsai, Yao-Hung Hubert and Bai, Shaojie and Liang, Paul Pu and Kolter, J. Zico and Morency, Louis-Philippe and Salakhutdinov, Ruslan},
  title     = {{Multimodal Transformer for Unaligned Multimodal Language Sequences}},
  booktitle = {Proceedings of the 57th Annual Meeting of the Association for Computational Linguistics},
  year      = {2019},
  pages     = {6558--6569},
  doi       = {10.18653/v1/P19-1656},
  url       = {https://aclanthology.org/P19-1656/}
}

@inproceedings{TFN,
  author    = {Zadeh, Amir and Chen, Minghai and Poria, Soujanya and Cambria, Erik and Morency, Louis-Philippe},
  title     = {{Tensor Fusion Network for Multimodal Sentiment Analysis}},
  booktitle = {Proceedings of the 2017 Conference on Empirical Methods in Natural Language Processing},
  year      = {2017},
  pages     = {1103--1114},
  doi       = {10.18653/v1/D17-1115},
  url       = {https://aclanthology.org/D17-1115/}
}

@inproceedings{LMF,
  author    = {Liu, Zhun and Shen, Ying and Lakshminarasimhan, Varun Bharadhwaj and Liang, Paul Pu and Zadeh, AmirAli Bagher and Morency, Louis-Philippe},
  title     = {{Efficient Low-Rank Multimodal Fusion with Modality-Specific Factors}},
  booktitle = {Proceedings of the 56th Annual Meeting of the Association for Computational Linguistics (Volume 1: Long Papers)},
  year      = {2018},
  pages     = {2247--2256},
  doi       = {10.18653/v1/P18-1209},
  url       = {https://aclanthology.org/P18-1209/}
}

@inproceedings{self-mm,
  author    = {Yu, Wenmeng and Xu, Hua and Yuan, Ziqi and Wu, Jiele},
  title     = {{Learning Modality-Specific Representations with Self-Supervised Multi-Task Learning for Multimodal Sentiment Analysis}},
  booktitle = {Proceedings of the AAAI Conference on Artificial Intelligence},
  volume    = {35},
  number    = {12},
  year      = {2021},
  pages     = {10790--10797},
  doi       = {10.1609/AAAI.V35I12.17289}
}

@article{Cmu-mosi,
  author  = {Zadeh, Amir and Zellers, Rowan and Pincus, Eli and Morency, Louis-Philippe},
  title   = {{Multimodal Sentiment Intensity Analysis in Videos: Facial Gestures and Verbal Messages}},
  journal = {IEEE Intelligent Systems},
  year    = {2016},
  volume  = {31},
  number  = {6},
  pages   = {82--88},
  doi     = {10.1109/MIS.2016.94}
}

@inproceedings{Cmu-mosei,
  author    = {Bagher Zadeh, AmirAli and Liang, Paul Pu and Poria, Soujanya and Cambria, Erik and Morency, Louis-Philippe},
  title     = {{Multimodal Language Analysis in the Wild: {CMU}-{MOSEI} Dataset and Interpretable Dynamic Fusion Graph}},
  booktitle = {Proceedings of the 56th Annual Meeting of the Association for Computational Linguistics (Volume 1: Long Papers)},
  year      = {2018},
  pages     = {2236--2246},
  doi       = {10.18653/v1/P18-1208},
  url       = {https://aclanthology.org/P18-1208/}
}

@article{iemocap,
  author  = {Busso, Carlos and Bulut, Murtaza and Lee, Chi-Chun and Kazemzadeh, Abe and Mower, Emily and Kim, Samuel and Chang, Jeannette N. and Lee, Sungbok and Narayanan, Shrikanth S.},
  title   = {{IEMOCAP: Interactive Emotional Dyadic Motion Capture Database}},
  journal = {Language Resources and Evaluation},
  year    = {2008},
  volume  = {42},
  number  = {4},
  pages   = {335--359},
  doi     = {10.1007/s10579-008-9076-6}
}

@inproceedings{sims,
  author    = {Yu, Wenmeng and Xu, Hua and Meng, Fanyang and Zhu, Yilin and Ma, Yixiao and Wu, Jiele and Zou, Jiyun and Yang, Kaicheng},
  title     = {{CH-SIMS: A Chinese Multimodal Sentiment Analysis Dataset with Fine-Grained Annotation of Modality}},
  booktitle = {Proceedings of the 58th Annual Meeting of the Association for Computational Linguistics},
  year      = {2020},
  pages     = {3718--3727},
  doi       = {10.18653/v1/2020.acl-main.343},
  url       = {https://aclanthology.org/2020.acl-main.343/}
}

@article{t-sne,
  author  = {van der Maaten, Laurens and Hinton, Geoffrey},
  title   = {{Visualizing Data Using t-SNE}},
  journal = {Journal of Machine Learning Research},
  year    = {2008},
  volume  = {9},
  number  = {86},
  pages   = {2579--2605},
  url     = {https://www.jmlr.org/papers/v9/vandermaaten08a.html}
}

@inproceedings{HRLF,
  author    = {Li, Mingcheng and Yang, Dingkang and Liu, Yang and Wang, Shunli and Chen, Jiawei and Wang, Shuaibing and Wei, Jinjie and Jiang, Yue and Xu, Qingyao and Hou, Xiaolu and Sun, Mingyang and Qian, Ziyun and Kou, Dongliang and Zhang, Lihua},
  title     = {{Toward Robust Incomplete Multimodal Sentiment Analysis via Hierarchical Representation Learning}},
  booktitle = {Advances in Neural Information Processing Systems},
  volume    = {37},
  year      = {2024},
  pages     = {28515--28536},
  doi       = {10.52202/079017-0895}
}

@inproceedings{DPCN,
  author    = {Hu, Zhangfeng and Zheng, Wenming and Zong, Yuan and Wei, Mengting and Jiang, Xingxun and Shi, Mengxin},
  title     = {{A Novel Decoupled Prototype Completion Network for Incomplete Multimodal Emotion Recognition}},
  booktitle = {2024 IEEE International Conference on Multimedia and Expo (ICME)},
  year      = {2024},
  pages     = {1--6},
  doi       = {10.1109/ICME57554.2024.10687673}
}

@inproceedings{LadderNetworks,
  author    = {Rasmus, Antti and Berglund, Mathias and Honkala, Mikko and Valpola, Harri and Raiko, Tapani},
  title     = {{Semi-Supervised Learning with Ladder Networks}},
  booktitle = {Advances in Neural Information Processing Systems},
  volume    = {28},
  year      = {2015},
  pages     = {3546--3554},
  url       = {https://proceedings.neurips.cc/paper/2015/hash/378a063b8fdb1db941e34f4bde584c7d-Abstract.html}
}

 

% \end{IEEEbiography}



\vfill

\end{document}